\documentclass[
 reprint,
superscriptaddress,
 amsmath,amssymb,
 aps,
 prxquantum,
 longbibliography,
]{revtex4-2}

\usepackage{graphicx}
\usepackage{dcolumn}
\usepackage{bm}
\usepackage{ulem}
\usepackage{float}
\usepackage[makeroom]{cancel}
\usepackage{dcolumn}
\usepackage{makecell}
\usepackage[utf8]{inputenc}
\usepackage[T1]{fontenc}
\usepackage{mathptmx}
\usepackage{commath}
\usepackage{multirow}
\usepackage{amsmath}

\usepackage{soul}
\usepackage{verbatim}
\usepackage{bm}
\usepackage{outlines}
\usepackage{siunitx}
\usepackage[space]{grffile}
\usepackage{upgreek}
\usepackage[dvipsnames]{xcolor}
\usepackage{xr}
\usepackage{color}
\usepackage{tikz}
\usepackage{mathtools}
\usepackage{amssymb,placeins}
\usepackage{hyperref}
\usepackage[table]{xcolor}
\usepackage{tabularx}
\usepackage{amsmath, mathtools, amssymb, dsfont, amsthm}
\usepackage{physics}
\usepackage{enumitem}
\usepackage{etoolbox}
\usepackage{placeins}

\newcommand{\ketg}{\left| g \right\rangle}
\newcommand{\kete}{\left| e \right\rangle}
\newcommand{\ketf}{\left| f \right\rangle}
\newcommand{\ketfzero}{\left| f, 0 \right\rangle}
\newcommand{\ketezero}{\left| e, 0 \right\rangle}
\newcommand{\ketp}{\left| + \right\rangle}
\newcommand{\ketfn}{\left| f, n \right\rangle}
\newcommand{\ketfm}{\left| f, m \right\rangle}

\newcommand{\ketem}{\left| e, m \right\rangle}
\newcommand{\tdelay}{t_{\mathrm{delay}}}
\newcommand{\lt}{\widetilde{L_1}}
\newcommand{\pL}{p_L}
\newcommand{\ns}{\mathrm{ns}}

\newcommand{\QuTech}{\affiliation{QuTech, Delft University of Technology, P.O. Box 5046, 2600 GA Delft, The Netherlands}}
\newcommand{\Kavli}{\affiliation{Kavli Institute of Nanoscience, Delft University of Technology, P.O. Box 5046, 2600 GA Delft, The Netherlands}}
\newcommand{\TNO}{\affiliation{Netherlands Organisation for Applied Scientific Research (TNO),P.O. Box 96864, 2509 JG The Hague, The Netherlands}}
\newcommand{\AM}{\affiliation{Delft Institute of Applied Mathematics, P.O. Box 5046,2600 GA Delft, Delft University of Technology, The Netherlands}}
\newcommand{\SRON}{\altaffiliation[Present address: ]{SRON Netherlands Institute for Space Research, Niels Bohrweg 4, 2333 CA Leiden, The Netherlands}}

\begin{document}
\preprint{APS/123-QED}

\title{Improved error correction with leakage reduction units built into qubit measurement in a superconducting quantum processor}

\author{Yuejie~Xin}\QuTech\Kavli
\author{Sean~L.~M.~van~der~Meer}\QuTech\Kavli
\author{Marc~Serra-Peralta}\QuTech\AM
\author{Tim~H.~F.~Vroomans}\QuTech\Kavli
\author{Matvey~Finkel}\QuTech\Kavli
\author{Hendrik~M.~Veen}\SRON\QuTech\Kavli
\author{Mark~W.~Beekman}\TNO
\author{Leonardo~DiCarlo}\email{Corresponding author: l.dicarlo@tudelft.nl}\QuTech\Kavli

\date{\today}

\begin{abstract}
Leakage to non-computational states is a source of correlated errors in both time and space that limits the effectiveness of quantum error correction (QEC) with superconducting circuits. We present and experimentally demonstrate a high-fidelity, leakage reduction unit (LRU) operating concurrently with transmon measurement without incurring time overhead. Adapted from double-drive reset of population (DDROP), the protocol utilizes simultaneous drives on the transmon and its readout resonator, leveraging the dispersive shift to create a directional process that returns the transmon to the computational subspace. The LRU achieves a 98.4\%  leakage removal fraction without compromising the computational-state assignment fidelity (99.2\%). We combine LRU-enhanced measurement and neural-network decoding to successfully suppress logical error rates in both memory and stability QEC experiments without any post-selection.
\end{abstract}

\maketitle

\section{Introduction}
Quantum error correction (QEC) provides a path towards fault-tolerant quantum computation by encoding information redundantly. Despite remarkable progress~\cite{Andersen20, Egan20, Ryan-Anderson21, Acharya23, Krinner22, Google25}, the performance of QEC is fundamentally limited by physical errors that corrupt quantum information. To achieve fault tolerance, it is crucial to understand and mitigate all significant error channels, especially those that are poorly handled by standard QEC codes. Among the various error channels, leakage of population out of the computational subspace is particularly damaging~\cite{Fowler13, Bultink20, Varbanov20}. For instance, in transmons, the most common superconducting qubit, the computational subspace is typically spanned by the ground ($\ketg$) and first excited ($\kete$) states, with excitation to the second excited state ($\ketf$) being the dominant type of leakage. This type of error poses a fundamental challenge to standard stabilizer codes like the surface code~\cite{Fowler12}. The power of these codes lies in their ability to discretize arbitrary physical errors that occur within the computational subspace into a set of correctable Pauli errors. However, leakage violates this core assumption because a leakage event cannot be mapped to the Pauli framework. Additionally, leakage can last for several QEC cycles, leading to correlated errors in time. Even worse, leakage can propagate through two-qubit gates, creating not only temporally but also spatially correlated errors that are highly detrimental to the logical performance~\cite{Aliferis07, Fowler13, Varbanov20, Mcewen21, Camps24}.

Various strategies have been developed to combat leakage. Hardware-level chip design can reduce it~\cite{Thorbeck24, Vazquez25}, while leakage-tolerant QEC codes have been proposed at the cost of increased number of qubits~\cite{Ghosh15, Eickbusch25}. Active methods like leakage reduction units (LRUs), which are designed to reset the qubit from the leakage state back into the computational subspace, offer a more direct solution~\cite{Mcewen21, Miao22, Marques23, Kim24, Chen24, Yang24, Lacroix25, Xiao25}. However, many LRU protocols require a dedicated time step within the QEC cycle. This adds idle time and can increase the overall physical error rate, potentially negating the benefits of leakage removal.

In this work, we address this challenge by experimentally demonstrating a high-fidelity and time-efficient LRU for superconducting qubits. Our protocol is inspired by DDROP~\cite{Geerlings12b}, which we adapt here for leakage removal from $\ketf$ to $\kete$. The key innovation is the concurrent operation of the LRU with the qubit measurement process, creating an LRU-enhanced measurement that has no time overhead compared to a standard readout. The protocol is broadly applicable, as its implementation only requires a dispersively coupled readout resonator, a nearly universal component in current superconducting quantum processors.

A high-fidelity three-level readout (3RO) can be achieved by the LRU-enhanced measurement, which provides the QEC decoder with significantly richer information than a conventional binary outcome. This has been shown to improve the logical performance~\cite{Ali24}. Specifically, a readout result corresponding to $\ketf$ heralds a leakage event at a known location (measure qubit) and time (QEC round). This transforms a non-Pauli error into a tractable, heralded error for the decoder. 
A key result of this work is a demonstration that QEC without any post-selection performs best when combining the LRU with 3RO.

The manuscript is organized as follows. We first detail the calibration and characterization of the LRU, quantifying its leakage removal efficiency and its minimal impact on readout fidelity. We then demonstrate its practical value by integrating it into two QEC benchmarks: a memory experiment of a distance-3 bit-flip repetition code and a 7-qubit stability experiment. In both cases, we show that combining the LRU with 3RO yields the highest logical performance among the four variants (2RO or 3RO with or without LRU) at all levels of controlled leakage injection on Controlled-$Z$ (CZ) gates.

\section{Implementation of the leakage reduction unit}
\begin{figure}[t]
\includegraphics[width=\columnwidth]{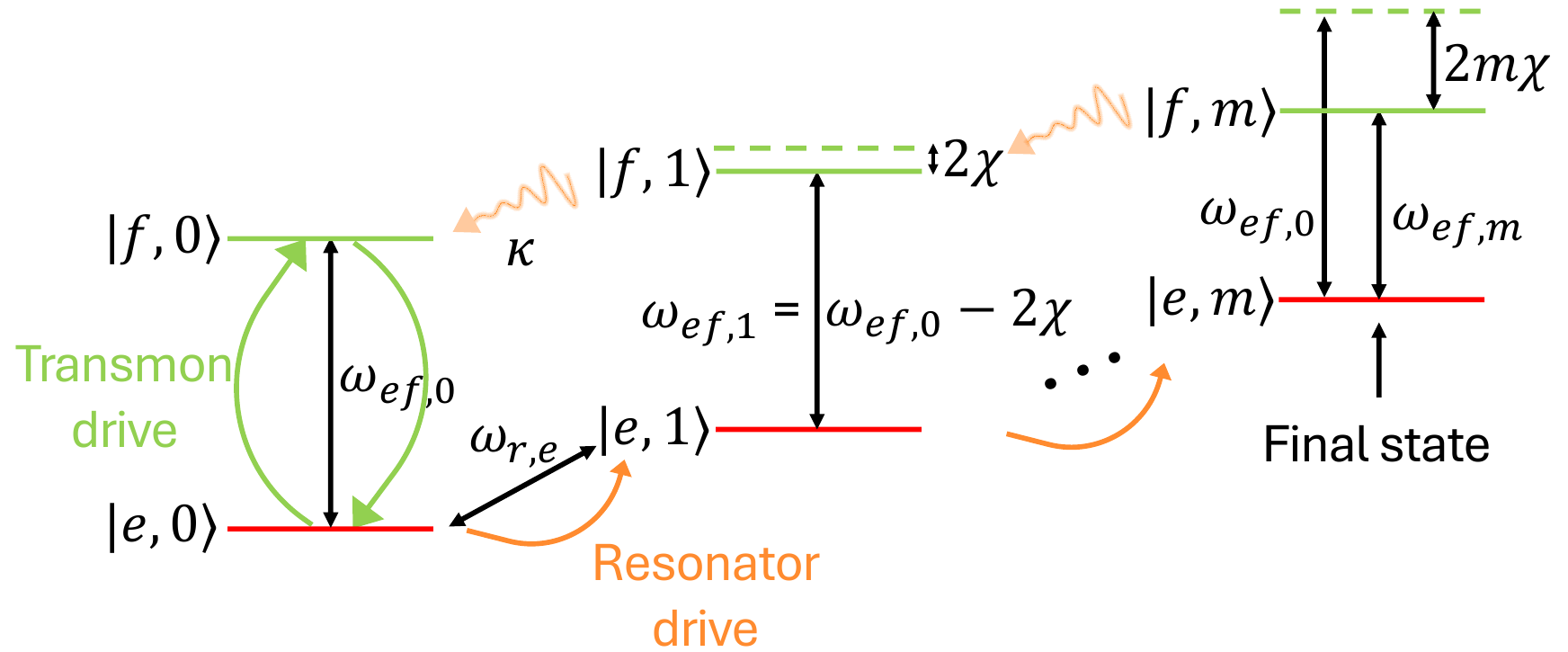}
\label{fig1}
\caption{
Simplified schematic of the leakage reduction unit (LRU) protocol. The energy levels of the coupled transmon-resonator system are shown, with states labeled as $|$Transmon, Resonator$\rangle$. The protocol utilizes two simultaneous drives: a transmon drive at the frequency of the $\kete\leftrightarrow\ketf$ transition when there is no photon in the resonator, $\omega_{ef,0}$, and a resonator drive at the resonator frequency when the transmon is in $\kete$, $\omega_{r,e}$.
The process begins with the system in an arbitrary leakage state $\ketfn$. The resonator population quickly decays to zero at a rate $\kappa$, resulting in the state $\ketfzero$. The transmon drive, which is resonant with the $\ketfzero\leftrightarrow\ketezero$ transition, then transfers the population to $\ketezero$. Subsequently, the resonator drive populates the resonator, moving the system to a final steady state, $\ketem$.
The directionality of this process is ensured by the dispersive shift, 2$\chi$. As photons accumulate in the resonator, the transmon transition frequency shifts. The original transmon drive at $\omega_{ef,0}$ becomes off-resonant with the $\ketem\leftrightarrow\ketfm$ transition by an amount 2$m\chi$, preventing the system from being re-excited to the $\ketf$ state manifold. Likewise, the resonator drive is off-resonant when the transmon is in $\ketf$, preventing photon accumulation. Once the drives are removed, the system quickly decays to $\ketezero$, effectively removing leakage from $\ketf$.
}
\end{figure}

The processor used in our experiments~\cite{Marques23} has a standard circuit quantum electrodynamics (cQED) architecture, with each transmon dispersively coupled to a dedicated readout resonator. The LRU protocol relies on a directional process for unconditional reset. The implementation, shown schematically in Fig.~1, is specifically tailored for leakage removal from $\ketf$ to $\kete$. For clarity and simplicity, this schematic omits the dedicated Purcell filter, which is present in our processor to protect each transmon from Purcell decay through the resonator. As we show below (Figs.~2 and~3), the LRU protocol is fully compatible with and effective in the presence of the filter.

The protocol creates a one-way path from the leakage state $\ketf$ to the computational state $\kete$ by applying simultaneous microwave drives to the transmon and the resonator. The core principle is to leverage the state-dependent frequencies of the transmon and resonator. A resonant transmon drive transfers population from $\ketfzero$ to $\ketezero$, while a second drive begins to populate the resonator, a process that is efficient only when the transmon is in $\kete$. This induced photon population is the key: it activates the dispersive shift, which detunes the transmon transition frequency. This detuning effectively locks the transmon in the computational subspace by making the reverse transition to $\ketf$ highly off-resonant.

Transitioning from the idealized model above to a realistic system introduces practical limitations that can affect performance due to finite linewidths. In a real system, an off-resonant drive can still induce unwanted transitions, which could reduce the leakage removal efficiency or even re-introduce leakage. This non-ideal effect can be mitigated by designing a system with a larger dispersive shift or decoherence time $T_2$ (detailed system parameters of our device can be found in Section I of~\cite{SOMDDLRU}). As will be shown later, tuning the drive parameters to find an optimal operating point also helps to reduce this problem.

A core advantage of driving the readout resonator is that measurement can be performed simultaneously with the leakage removal. This effectively creates an LRU-enhanced measurement with no time overhead. Furthermore, the transmon drive is designed to only induce the transition between $\kete$ and $\ketf$, ideally leaving the ground state $\ketg$ unaffected. We have experimentally confirmed that the assignment fidelity between $\ketg$ and $\kete$ remains uncompromised during this process (see Fig.~3). By introducing a delay, $\tdelay$, between the resonator and transmon drives, a distinct readout signal for $\ketf$ can also be obtained (see Fig.~2 for details).

For calibration, both assignment fidelity and leakage removal efficiency are optimized  using the circuit shown in Fig.~2(a). The transmon drive pulse has the sine-squared envelope proposed in Ref.~\cite{Battistel21}, with $10~\ns$ rise and fall times, while the resonator drive is a square pulse. The total duration for the LRU-enhanced measurement is $500~\ns$ ($380~\ns$ pulses followed by $120~\ns$ for photon depletion), which typical for a standard measurement in our system. The final population tomography [quantified by a 3$\times$3 matrix as shown in Fig.~3(e,f)] is corrected for readout errors using an iterative Bayesian unfolding method~\cite{ibu}, with a reference matrix for unfolding averaged over 10,000 shots.

\begin{figure}[t]
\includegraphics[width=\columnwidth]{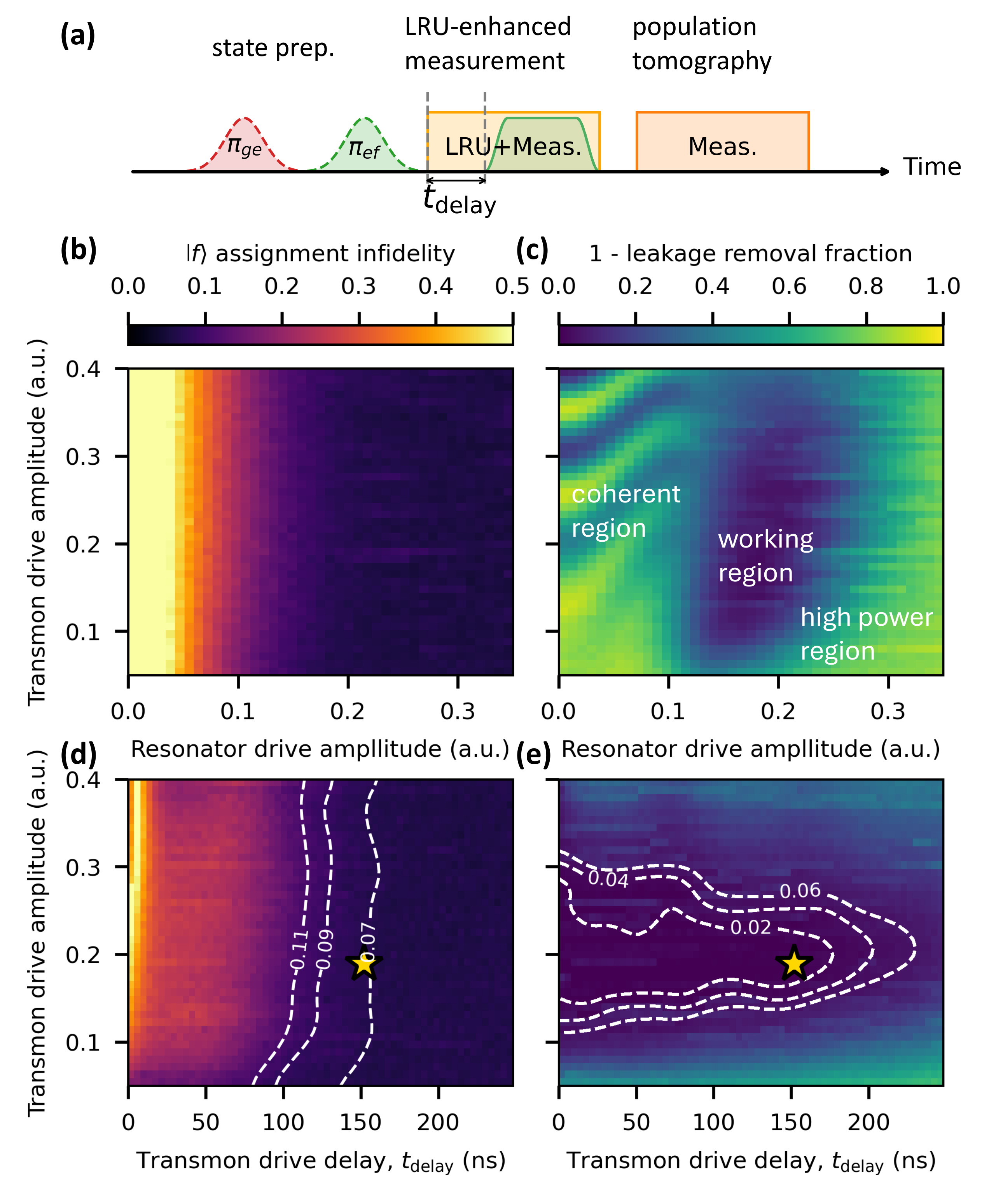}
\label{fig2}
\caption{
Calibration of the LRU pulse.
(a) The experimental pulse sequence used for calibration. The transmon is first prepared in the $\ketg$, $\kete$, or $\ketf$, followed by the LRU-enhanced measurement. A delay, $\tdelay$, is introduced between the resonator and transmon drives to enable three-level readout. Finally, population tomography is performed using a standard measurement with readout errors corrected.
(b--e) Performance versus drive parameters. Each point averages over 4,000 shots.
In (c), three distinct behaviors are observed: a coherent region at low resonator power where Rabi oscillations dominate; a central working region where balanced drives enable the intended LRU process; and a high power region where the strong resonator drive populates the resonator with photons exceeding the critical photon number of the dispersive regime.
In (d, e), the trade-off between assignment fidelity and leakage removal is explored. In (d), the assignment fidelity increases with $\tdelay$ before being limited due to transmon relaxation. In (e), the residual $\ketf$ population increases as $\tdelay$ gets longer; it rises slowly at first while photons build up in the resonator, then rises more quickly as the reduced time for the removal process becomes the dominant factor. The white dashed lines indicate contours for infidelity (0.07) and residual population (0.02). The yellow star marks the final choice of parameters.
}
\end{figure}

Due to the coupled nature of the drive parameters, the covariance matrix adaptation evolution strategy (CMA-ES) algorithm~\cite{cma} is used to find a global optimum. We use the geometric mean of assignment fidelity and leakage removal as the cost function, while the parameters to optimize include the frequencies and amplitudes of the two drives, as well as $\tdelay$. To illustrate the underlying physics, Fig.~2(b--e) shows 2D sweeps of the key parameters, with all other parameters held at their optimized values.

There are three distinct physical regimes that depend on both drive amplitudes, as highlighted in the landscape of Fig.~2(c). The performance is quantified by the leakage removal fraction, defined as 1 - $P(f|f)$ and obtained from the population tomography, where $P(f|f)$ is the probability of both start and end states being $\ketf$. The first regime, the coherent region, occurs when the resonator drive is too weak, and an insufficient number of photons populates the resonator. The resulting dispersive shift is not large enough to prevent the transmon drive from causing off-resonant transitions, leading to coherent Rabi oscillations between $\kete$ and $\ketf$ rather than a directional reset. This effect can also be quantitively described by the state jump rate $\gamma_{\mathrm{jump}}=\Omega_R^2/2(\gamma_2+\Gamma_d)$, where $\Omega_R$ is the Rabi drive rate, $\gamma_2$ is the qubit dephasing rate and $\Gamma_d$ is the measurement-induced dephasing rate~\cite{Gambetta08}. Conversely, if the resonator drive is too strong (the high power region), the number of photons exceeds the critical photon number, and the system no longer operates in the dispersive regime, breaking a fundamental assumption of the protocol. Between these two extremes lies a broad working region where the drives are balanced, and the  LRU process proceeds with high efficiency. The large area of this working region demonstrates the robustness of the protocol against variations in drive amplitudes.

One of the key features of the protocol is the ability to enable a distinct readout of $\ketf$ without much cost on time and leakage removal. As shown in the circuit in Fig.~2(a), this is achieved with finite $\tdelay$. This delay creates a pure readout window where the resonator response is sensitive to the transmon state before the leakage removal begins. Note that while the delay necessarily reduces the available time for the subsequent leakage removal, it is not simply a penalty. Building up photons in the resonator before applying the transmon drive helps to reduce leakage induced by the transmon drive itself and can speed up the subsequent reset process. Varying $\tdelay$ reveals a performance trade-off, as illustrated in Fig.~2(d-e). We define the $\ketf$ assignment infidelity [Fig.~2(b, d)] as $1-P(2|f)$, where $P(2|f)$ is the probability of the LRU-enhanced measurement outputting 2 when the transmon is prepared in $\ketf$. In Fig.~2(d), increasing $\tdelay$ initially improves the $\ketf$ assignment fidelity by allowing more time for pure readout. This improvement continues until the fidelity becomes limited by the transmon natural relaxation time. In Fig.~2(e), the residual $\ketf$ population increases slowly for shorter $\tdelay$, as there are other limiting factors. Then the residual population rises more quickly for longer $\tdelay$ as the insufficient pulse duration dominates. Effectively, we can gain the significant benefit of a high-fidelity $\ketf$ state readout without a major compromise in leakage removal performance. The chosen operating point, marked by the yellow star, is where these factors are balanced, achieving a near-saturated assignment fidelity while maintaining a high leakage removal fraction.

\begin{figure}[t]
\includegraphics[width=\columnwidth]{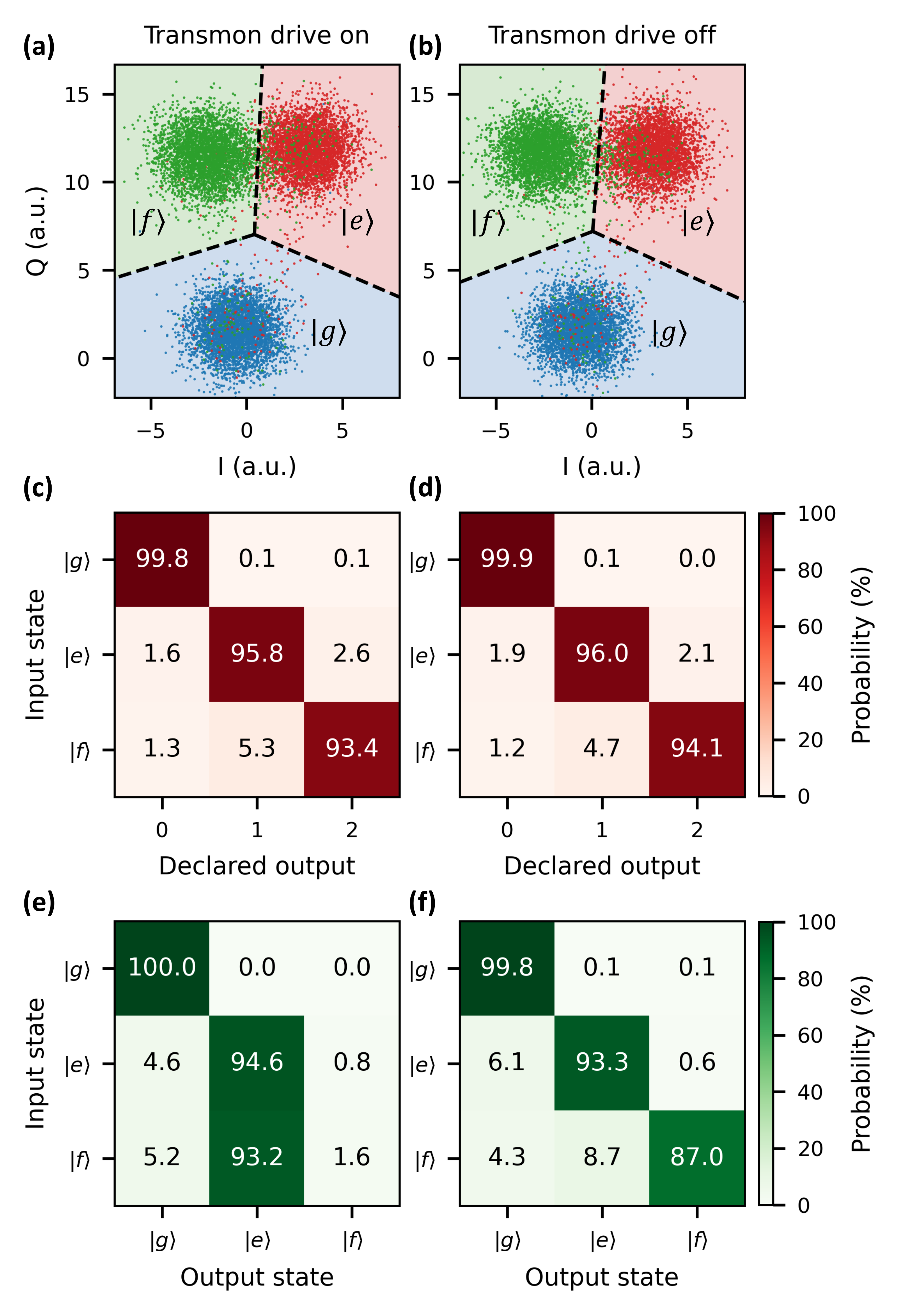}
\label{fig3}
\caption{
Performance comparison of LRU-enhanced and standard measurements. The LRU-enhanced measurement (left column) is compared to a standard measurement (right column), which is implemented by turning off the transmon drive while keeping all other parameters identical.
(a, b) Single-shot I-Q data with 4,000 shots for each state.
(c, d) Readout assignment matrices, showing the probability of assigning a measurement outcome given a prepared state. The average assignment fidelities are 98.2\% (LRU) and 98.3\% (standard).
(e, f) Population transfer matrices, showing the probability of the transmon being in a final state given its initial state. The matrix in (e) demonstrates a leakage removal fraction of 98.4\%.
}
\end{figure}

With the LRU protocol calibrated, its performance is now directly compared against a standard measurement in Fig.~3. To ensure a fair comparison, the standard measurement is implemented by simply turning off the transmon drive of the LRU-enhanced measurement, while keeping all other parameters identical. The single-shot readout data, plotted on the I-Q plane in Fig.~3(a, b), shows well-separated state clusters for both measurement types. A more quantitative analysis is provided by the readout assignment matrices in Fig.~3(c, d). The average 3RO assignment fidelities are nearly identical at 98.2\% for the LRU-enhanced measurement and 98.3\% for the standard one. To be specific, the assignment fidelities between $\ketg$-$\kete$ and $\ketg$-$\ketf$ are unchanged. This confirms that the LRU does not compromise the overall ability to distinguish the computational states. The fidelity between $\kete$ and $\ketf$ is slightly lower with the LRU, consistent with the trade-off discussed in Fig.~2. The LRU performance is revealed in the population transfer matrices [Fig.~3(e, f)], which are extracted from the final readout-error-corrected population tomography [Fig.~2(a)]. These matrices show the statistics of the final state of the transmon after the measurement. The off-diagonal terms in the lower-left of both matrices, such as the probability $P(g|e)$, are primarily due to relaxation during the measurement window. The crucial result is in the bottom row of Fig.~3(e), which shows that if the transmon starts in $\ketf$, the LRU process actively returns it to $\kete$. This corresponds to a leakage removal fraction of 98.4\%. In contrast, the matrix for the standard measurement in Fig.~3(f) shows a high $P(f|f)$, indicating that the leakage state persists as expected.

\section{LRU in quantum error correction}
Having established and calibrated the LRU-enhanced measurement, we explore its benefit in the context of QEC. We perform two experiments that benchmark the basic tasks in fault-tolerant quantum computation: the memory experiment~\cite{Google21, Zhao22b, Krinner22, Sundaresan23, Acharya23, Google25}, which tests the preservation of a logical observable through time, and the stability experiment~\cite{Gidney22, Caune24, Harper25}, which tests the movement of a logical observable through space, as in lattice surgery~\cite{Horsman12}. Leakage is a particularly damaging error source in these two contexts, as a leaked transmon can induce correlated errors in subsequent rounds. By actively returning leaked transmons to the computational subspace, our LRU protocol is expected to suppress the impact from these error channels and thereby improve the logical performance.

\subsection{Memory experiment}
The memory experiment is performed using a distance-3 bit-flip repetition code, which we call Repetition-5. As shown in the device schematic in Fig.~4(a), the logical qubit is encoded in three data qubits ($D_7, D_8, D_9$), while two measure qubits ($Z_3, Z_4$) are used to perform parity checks (same device as in Ref.~\cite{Marques23}, but with $X$ and $Z$ labels swapped). As the experimental circuit in Fig.~4(a) shows, after initializing the data qubits into one of the basis states $|i, j, k\rangle, i, j, k \in \{0, 1\}$, the system undergoes $R$ rounds of stabilizer measurements. At the end of each round, $X$ gates are applied to data qubits during measure qubit readout to make state-dependent errors (mainly relaxation here) more state-independent. These $X$ gates are replaced by two-level readout (2RO) on data qubits in the last round [not shown in Fig.~4(a) for simplicity]. The error information can be extracted from readout outcomes of measure qubits. A parity is defined by the difference between two consecutive measure qubit outcomes in our case, as we do not reset the measure qubits each round. The parity of the first (last) round is directly computed from data qubit initialization (measurements). A change in the parity relative to the previous round signals a detection event, indicating that an error may have occurred in the interim. The space-time history of these detection events, together with extra information (leakage flags in our case) are fed to a decoder. We use a neural network decoder as described in Sections III and IV of~\cite{SOMDDLRU}. To avoid the neural network decoder learning unwanted information, random logical bit flips (dashed $X$ gates) are applied at the start of each QEC round, as shown in Fig.~4(a).

\begin{figure}[t]
\includegraphics[width=\columnwidth]{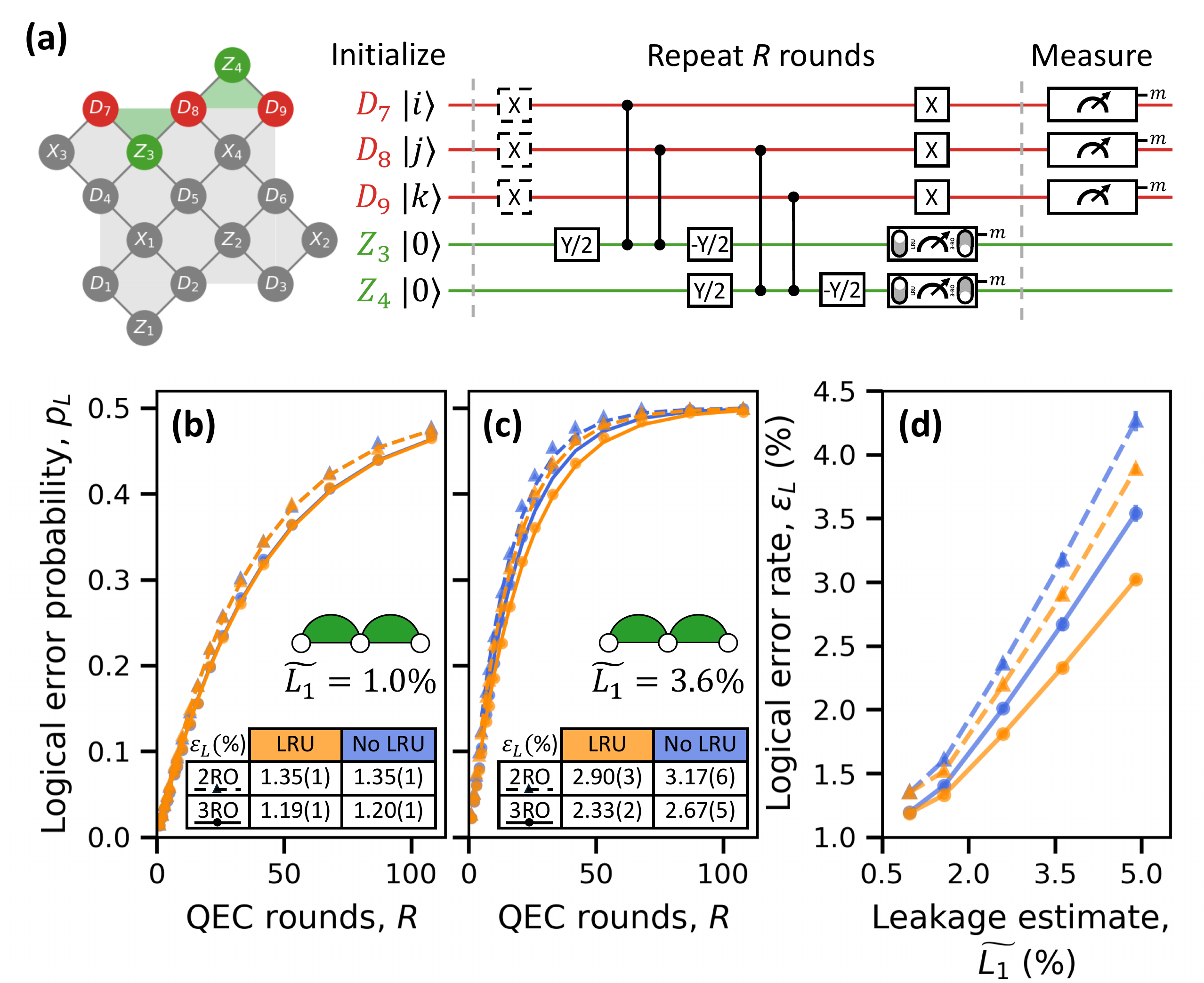}
\label{fig4}
\caption{
Memory experiment of a distance-3 bit-flip repetition code.
(a) Schematic showing the five qubits used, with data qubits in red and measure qubits in green. Quantum circuit consisting of qubit initialization, $R$ rounds of stabilizer measurements, and final measurement.
(b, c) Logical error probability ($\pL$) as a function of the number of QEC rounds ($R$) at baseline leakage ($\lt=1.0\%$) and intermediate leakage ($\lt=3.6\%$). Performance is compared for protocols with LRU (orange) and without (No LRU, blue), using either a three-level (3RO, solid lines, circles) or two-level (2RO, dashed lines, triangles) readout on the measure qubits. Results are averaged over all basis states $|i, j, k\rangle$.
(d) Extracted logical error rate per round ($\varepsilon_L$) as a function of the estimated leakage rate ($\lt$). The combination of LRU with 3RO consistently provides the lowest error rate, with the advantage growing as leakage increases.
}
\end{figure}

The goal of the decoder is to use the measured information to determine a logical correction. By repeating the experiment for a fixed number of rounds $R$, the logical error probability $\pL(R)$ is measured. The key figure of merit is the logical error rate per round, $\varepsilon_L$, which is extracted by fitting $\pL(R)=\frac{1}{2}-\frac{1}{2}(1-2\varepsilon_L)^{R+R_0}$, where $R_0$ is a constant~\cite{Obrien17}. To improve the extraction of $\varepsilon_L$, we fix $R_0=0$ and fit starting from round 5.

To study the effects of leakage, we controllably sweep the leakage rate of the CZ gates [Fig.~4(b--d)]. The leakage rate is swept by adjusting control parameters along the contour of $180^\circ$ conditional phase. As a proxy for leakage $L_1$, we use the upper bound $\lt$ obtained from conditional oscillation experiments~\cite{Rol19} (see Section II of~\cite{SOMDDLRU} for details). In all CZ gates, measure qubits are the ones prone to leak as they are the higher-frequency transmons of every pair.

Without intentional leakage injection [Fig.~4(b), at baseline $\lt = 1.0\%$] using 3RO over 2RO gives the most important performance improvement, reducing the logical error rate by 11.1\%.  (Note that 2RO is achieved by actually doing a 3RO and mapping the outcome 2 to 1.) Incorporating the LRU shows benefit at all other levels of injected leakage, for both 2RO and 3RO, as shown specifically for $\lt = 3.6\%$ in Fig.~4(c). The full leakage sweep in Fig.~4(d) shows that combining LRU and 3RO universally gives the best results. Further research is required to see if this holds at larger code distances and for other codes. In addition to investigating leakage during CZ gates in the experiment, we have used simulation to explore the impact of other leakage sources~(Section VI of~\cite{SOMDDLRU}).

\subsection{Stability experiment}
The stability experiment, which we call Stability-7, is performed with a 1D chain of four $Z$-type stabilizers involving seven qubits [Fig.~5(a)].  The experiment begins by initializing the four measure qubits in one of classical bitstrings $|i, j, k, l\rangle, i, j, k, l \in \{0, 1\}$, and the three data qubits in $\ketp$. The circuit is similar to that of Repetition-5; however, there is no need for random flips on data qubits as the logical observable is now defined by measure qubits instead of data qubits, and data qubit measurements are not used in the decoding. The logical observable of a stability experiment is defined as the product of the four measure qubit readout outcomes in a given round; here, we choose the last odd round. A logical error occurs if this measured product, after correction by the decoder, is different from the known fixed product (0 or 1 depending on the initial measure qubit states). Stability experiments can be viewed as space-time rotated memory experiments~\cite{Gidney22}, where increasing $R$ exponentially reduces $\pL$, analogous to increasing the code distance in a memory experiment. We fit the data to the model $\pL = A e^{-\gamma R}$, with error-suppression factor $\gamma$ and scaling factor $A$.  As the decay of $\pL$ is not purely exponential in the first few rounds (a point we address later in this section), we perform fits starting from round 10.  In this experiment, we  inject leakage only on the CZ gates between measure and data qubits in which the measure qubit is higher frequency and thus the one susceptible to leakage (all CZ gates except $D_5$-$Z_2$ and $D_5$-$Z_3$).

\begin{figure}[t]
\includegraphics[width=\columnwidth]{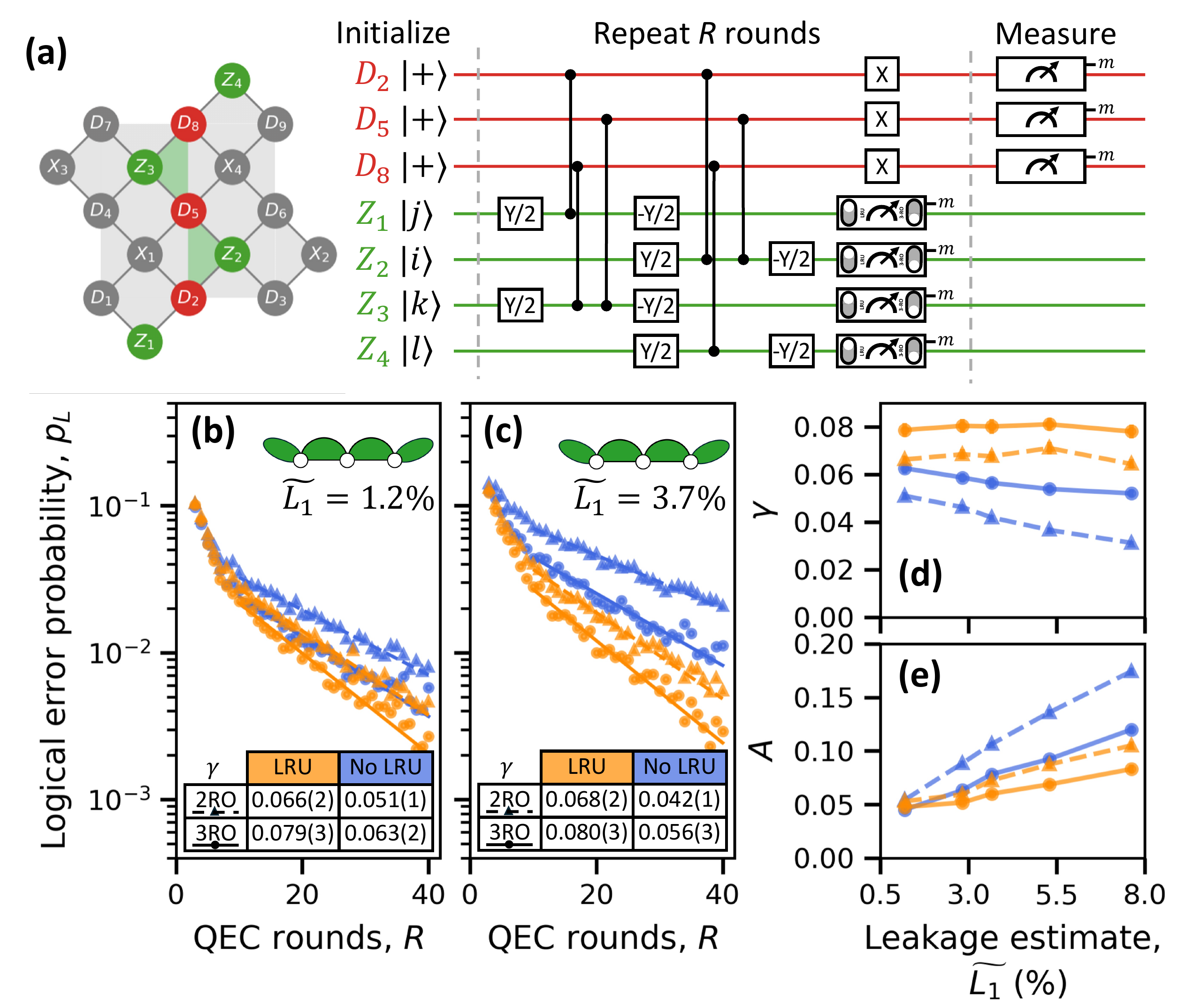}
\label{fig5}
\caption{Stability-7 experiment.
(a) Schematic of the seven qubits used and quantum circuit. Device and conventions are the same as in Fig.~4. CZ gates with narrower horizontal spacing between them are executed simultaneously.
(b, c) Logical error probability ($\pL$) versus QEC rounds ($R$) at baseline leakage ($\lt=1.2\%$) and intermediate leakage ($\lt=3.7\%$). The legend is the same as in Fig.~4. The data (scatters) is fit to $\pL = A e^{-\gamma R}$ starting from round 10 (solid lines).
(d, e) Extracted fitting parameters $\gamma$ (higher is better) and $A$ (lower is better) as a function of the leakage estimate $\lt$. 
With the LRU, the error suppression factor $\gamma$ remains stable against increasing leakage. 
The combination of LRU with 3RO consistently provides the best performance for both $\gamma$ and $A$.
}
\end{figure}

In Stability-7, using LRU and 3RO both in isolation and especially in combination gives substantial performance improvements even at the baseline leakage rate [Fig.~5(b)]. Compared to 2RO without LRU, $\gamma$ increases by $29\%$, $23\%$, and $54\%$ for 2RO with LRU, 3RO without LRU, and 3RO with LRU, respectively. We observe significant improvements even at baseline because stability experiments are more susceptible to time-like errors than memory experiments. Leakage is a time-correlated error and can be long lived, making it especially damaging if not dealt width. The improvement is even more significant at increasing levels of injected leakage.
For example, at $\lt=3.7\%$ [Fig.~5(c)], $\gamma$ increases by $61\%$, $33\%$, and $90\%$, respectively.

The non-exponential curvature of $p_L$ at low rounds in Fig.~5(b,c) warrants further discussion as it is not expected with LRU.
From simulation, we believe that this curvature is not caused by insufficient leakage removal by the LRU, nor by leakage on the central data qubit ($D_5$) which does not have an LRU.
We hypothesize that it may be due to leakage to higher excited states beyond $\ketf$.

The central finding is that the LRU protocol protects the logical observable against leakage, maintaining a nearly constant $\gamma$ even as the leakage rate increases [Fig.~5(d)].  Without leakage removal, $\gamma$ degrades significantly as the leakage rate increases. This demonstrates that the LRU is highly effective at reducing the long-lived time correlations from leakage events, which damages the logical performance of the stability experiment. Across the studied leakage rates, the combination of the 3RO with LRU consistently provides the best performance. This is especially valuable for lattice surgery and logical qubit mobility, because using LRUs with 3RO can potentially reduce the number of rounds required to achieve a certain $\pL$.

\section{Conclusion and outlook}
We have presented the mechanism, calibration, and experimental validation of a high-fidelity, zero-overhead LRU for superconducting qubits. By integrating the reset mechanism directly into the measurement pulse, we achieve a $98.4\%$ leakage removal fraction from $\ketf$ and 2RO (3RO) with average assignment fidelity of $99.2\%$ $(98.2\%)$. We demonstrated the practical value of this LRU-enhanced measurement by integrating it into two QEC benchmarks. In the Repetition-5 memory experiment, the logical error rate is most reduced by using 3RO, while adding the LRU helps further at finite injected leakage. In the Stability-7 experiment, the LRU successfully mitigates the propagation of leakage-induced correlated errors, evidenced by an increase in the error suppression factor even at baseline leakage,  and its stabilization against all levels of injected leakage. Throughout these experiments, we consistently found that leveraging the 3RO capability of the LRU provides a significant, additional benefit to the decoder, showcasing the power of combining active error removal with richer error syndrome information.

Our work establishes a practical and highly effective method for combatting one of the most detrimental error channels in superconducting quantum processors. The zero-overhead nature of this technique makes it a valuable tool for improving the performance and scalability of near-term QEC implementations. Future work could focus on deploying this LRU-enhanced measurement across larger-scale QEC codes. Microwave crosstalk may play a role in parallel LRUs, which we have not fully characterized. Besides, carefully designing system parameters to improve the leakage removal efficiency is also necessary for lower overall physical error schemes. Furthermore, it can also be combined with other data-qubit LRUs~\cite{Mcewen21, Marques23, Kim24, Chen24, Yang24, Lacroix25, Xiao25} during measurements to mitigate leakage on data qubits.

\begin{acknowledgments}
We thank Barbara Terhal, Mackenzie Shaw, and Yuli Nazarov for discussions, Nadia Haider for help in chip design, and Santiago Vallés-Sanclemente, Qblox and Orange Quantum System for technical contributions. This research is supported by the European Union Flagship on Quantum Technology (OpenSuperQplus100, no. 101113946), the Dutch National Growth Fund (KAT-1 and DiagnostiQ), Intel Corporation, and the Dutch Ministry of Economic Affairs (TKI).
\end{acknowledgments}

\section*{Author contributions}
YX conceptualized the QEC experiments. LDC conceptualized the LRU. YX performed the experiment with contributions from SLMM and THFV. YX performed the data analysis. MSP performed the NN decoding. MSP provided theory support for QEC experiments. MB and LDC designed the device. MF and HMV fabricated the device. YX wrote the manuscript with contributions from MSP, and feedback from SLMM and LDC. LDC supervised the project.

\section*{Conflicts of interests}
The authors declare no conflict of interests.

\section*{Data availability}
The data for all figures in main text and supplement are available at
\verb"https://doi.org/10.4121/"\\
\verb"902d7f9e-38bf-48a2-a2f7-52bbf7aeeedf".
Scripts for training the NNs are available at
\verb"https://github.com/MarcSerraPeralta/"
\verb"data-lru-integrated-with-measurement-for-qec".
Requests for additional materials should be addressed to LDC.

\clearpage

\renewcommand{\theequation}{S\arabic{equation}}
\renewcommand{\thefigure}{S\arabic{figure}}
\renewcommand{\thetable}{S\arabic{table}}
\renewcommand{\bibnumfmt}[1]{[S#1]}
\renewcommand{\citenumfont}[1]{S#1}
\setcounter{figure}{0}
\setcounter{equation}{0}
\setcounter{table}{0}
\setcounter{section}{0}

\onecolumngrid
\section*{Supplementary material for "Improved error correction with leakage reduction units built into qubit measurement in a superconducting quantum processor"}
\FloatBarrier
\twocolumngrid

This supplementary material provides additional information supporting the claims and figures of the main text.

\section{Device Metric}
Detailed device metrics are listed in Table S1. Note that the LRU-enhanced measurements in QEC experiments are calibrated with longer durations than in Fig. 2, due to parameter difference for different transmons.

\begin{table}[t]
\caption{Device metrics}
\begin{ruledtabular}
\begin{tabular}{lc}
Metric in LRU calibration & $Z_3$ 				\\
\colrule
Transmon frequency, $\omega_{ge}/2\pi$ (GHz) & 5.941 \\
Anharmonicity, $\alpha/2\pi$ (MHz) & 306 \\
Resonator frequency, $\omega_{\mathrm{r}, g}/2\pi$ (GHz) & 7.314 \\
Purcell frequency, $\omega_{\mathrm{p}, g}/2\pi$ (GHz) & 7.309 \\
Coupling strength R-P, $J/2\pi$ (MHz) & 9.3 (1) \\
Total dispersive shift, $2\chi$ (MHz) & 10.8 \\
Purcell linewidth, $\kappa/2\pi$ (MHz) & 24.8 (7) \\
LRU drive frequency on resonator (GHz) & 7.303 \\
LRU drive frequency on transmon (GHz) & 5.613 \\
Relaxation time, $T_1$ ($\mu$s) & 11 \\ 
Ramsey dephasing time, $T_2^*$ ($\mu$s) & 13 \\ 
\colrule
Metric in memory experiments & Average \\
\colrule
$\ketf$ assignment fidelity of measure qubit with LRU (\%) & 93 (2) \\
$\ketf$ assignment fidelity of measure qubit w/o LRU (\%) & 94 (3)\\
$\ketg$--$\kete$ assignment fidelity of measure qubit (\%) & 98.6 (1)\\
LRU leakage removal fraction (\%) & 95 (2) \\
$T_1$ of data qubit ($\mu$s) & 20 (9) \\ 
$T_1$ of measure qubit ($\mu$s) & 32 (8) \\ 
$T_2^*$ of measure qubit ($\mu$s) & 22 (9) \\ 
\colrule
Metric in stability experiments & Average \\
\colrule
$\ketf$ assignment fidelity of measure qubit with LRU (\%) & 93 (2)\\
$\ketf$ assignment fidelity of measure qubit w/o LRU (\%) & 94 (2)\\
$\ketg$--$\kete$ assignment fidelity of measure qubit (\%) & 98.3 (2)\\
LRU leakage removal fraction (\%) & 94.2 (2) \\
$T_1$ of data qubit ($\mu$s) & 13 (1) \\ 
$T_1$ of measure qubit ($\mu$s) & 16 (3) \\ 
$T_2^*$ of measure qubit ($\mu$s) & 23 (10) \\ 
\colrule
Operation in QEC experiments & Duration (ns) \\
\colrule
Single-qubit gate & 20 \\
Two-qubit gate & 60 \\
LRU-enhanced measurement & 700 \\
\end{tabular}
\end{ruledtabular}
\end{table}

\section{Sweeping and estimating CZ gate leakage}
\begin{figure}[t]
\includegraphics[width=\columnwidth]{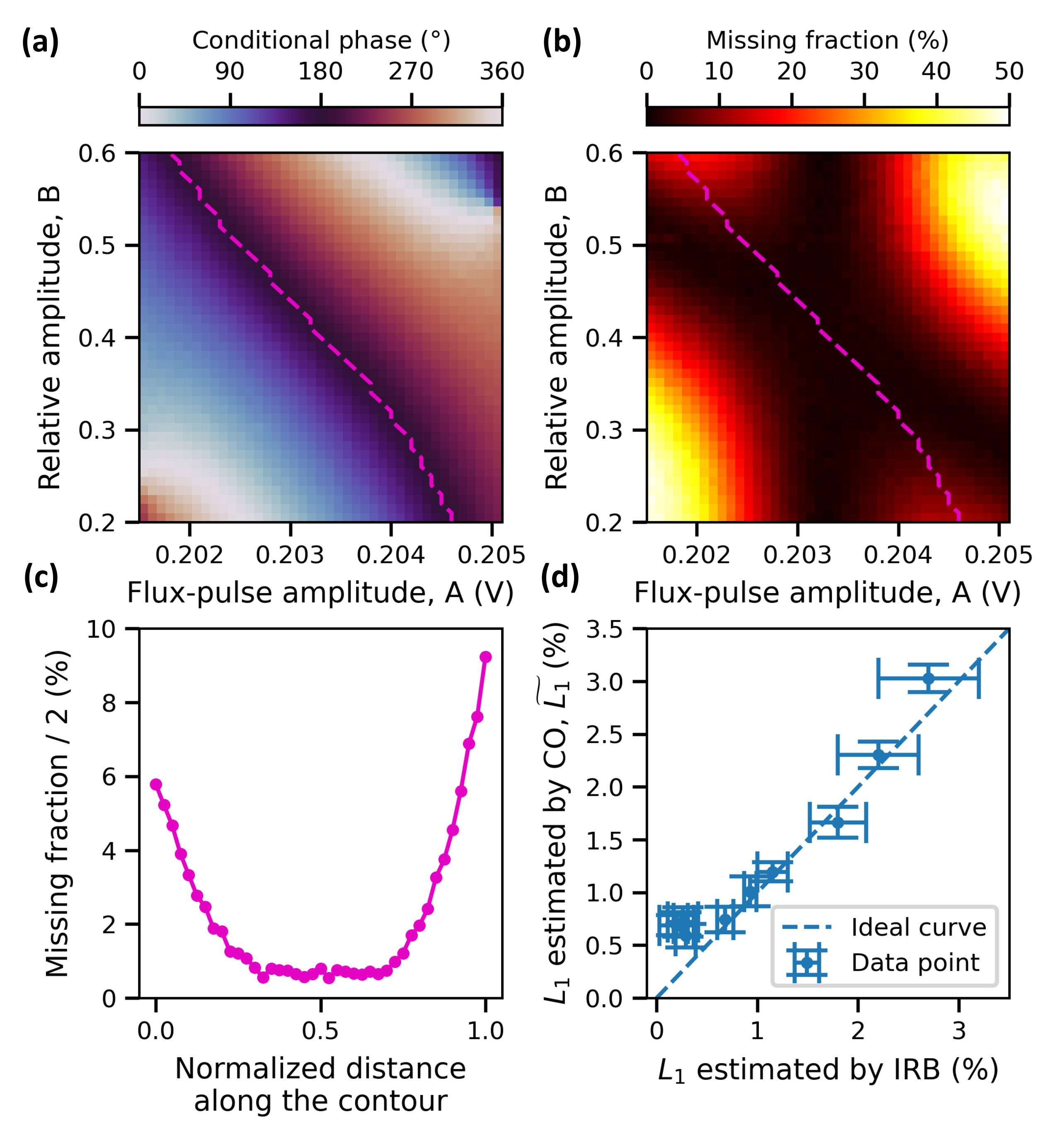}
\caption{Calibration and sweeping of CZ gate leakage.
(a) Conditional phase landscape as a function of flux-pulse amplitude ($A$) and relative amplitude ($B$), as defined in Ref.~\cite{Negirneac21}. The dashed pink line highlights the 180$^\circ$ contour. Each point is extracted from a 19-point conditional oscillation (CO) experiment with 1024 shots per point. The qubit pair is $X_2$--$D_2$.
(b) Missing fraction, $M$ measured across the same parameter landscape as in (a).
(c) Leakage estimate, taken as $M$ / 2, plotted along the 180$^\circ$ contour. This demonstrates how leakage can be swept while maintaining the correct conditional phase.
(d) Comparison of the leakage estimate derived from CO ($\widetilde{L_1}$) against the leakage measured by Interleaved Randomized Benchmarking (IRB). Ideally $\widetilde{L_1}=L_1$ (dashed curve). In reality, $\widetilde{L_1}$ provides a upper bound for $L_1$ that closely tracks the IRB value at higher leakage rates. For this comparison, CO experiments used 2$^{14}$ shots per point, and IRB used 200 seeds with 2$^{10}$ shots each. 
}
\label{som_cz}
\end{figure}

To investigate the impact of leakage on logical qubit performance, a method is required to controllably sweep the leakage rate ($L_1$,~\cite{Wood18}) of CZ gates while maintaining their 180$^\circ$ conditional phase. Following the procedure in Ref.~\cite{Negirneac21}, we use conditional oscillation (CO) experiments to map both the conditional phase [Fig.~\ref{som_cz}(a)] and the missing fraction $M$ [Fig.~\ref{som_cz}(b)] as a function of the flux-pulse amplitude $A$ and relative amplitude $B$. By selecting ($A$, $B$) parameter along the 180$^\circ$ conditional phase contour [pink dashed curves in Fig.~\ref{som_cz}(a--c)], CZ gates with the correct conditional phase but controlled leakage can be implemented.

We use $\widetilde{L_1}=M/2$ as a proxy for leakage as $\widetilde{L_1} \geq L_1$, with equality in the absence of decoherence.  We compare this CO-based estimation of leakage against $L_1$ extracted from Interleaved Randomized Benchmarking (IRB) with modifications to quantify leakage~\cite{Wood18} [Fig.~\ref{som_cz}(d)]. The overestimation of $L_1$ by $\widetilde{L_1}$  exceeds error bars only at the very lowest levels of leakage. Given that CO experiments are substantially faster than IRB, we use $\widetilde{L_1}$ as a proxy for $L_1$ in the QEC experiments.

\section{Details of the neural network decoder}
In this section, we provide the details of the neural network (NN) decoding results in Figs.~4 and~5. These include the architecture, training, inputs, ensembling of the NN, and definition of logical observables. 

The training of NNs follows Refs.~\cite{som_Ali24} and~\cite{Varbanov25}: in the sections below we only report the parameters that we have explicitly changed. We refer to our repository~\cite{data-project} for the complete training and parameter description. 

For both Repetition-5 and Stability-7 experiments, there are three datasets: training, validation, and testing. The NNs have been trained using the training dataset while monitoring their performance with the validation dataset to early stop and to check for overfitting, among others. The reported performance is obtained from decoding the testing dataset, which has not been previously seen by the NN. We use the same experimental dataset for the two-level readout (2RO) and three-level readout (3RO) cases in Figs.~4 and~5 by processing the I-Q voltages of the measurements using both a 3RO and a 2RO classifier. Note that our 2RO classifier corresponds to our 3RO classifier but outputs $m$=1 when the ternary measurement outcomes, denoted as $\Tilde{m}$, has $\Tilde{m}=2$. To avoid human bias towards favoring the performance of the LRU with 3RO when selecting the NN sizes and training hyperparameters, we have optimized them using only the dataset with 2RO and without LRUs. The best values found are then used across all NNs. Such a choice may not be optimal for decoding the experiments using the LRU with 3RO, but it ensures a fair comparison between the different settings.

\subsection{NN architecture, training hyperparameters, and inputs} \label{sec:NN-para}

\begin{figure}[t]
    \centering
    \includegraphics[width=65mm]{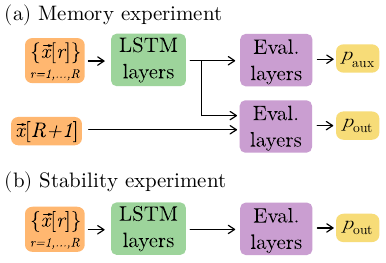}
    \caption{Our neural network (NN) architecture for (a) memory experiments and (b) stability experiments. The inputs $\Vec{x}[r]$ contain the information described in Section~\ref{sec:nn_inputs} for QEC round $r$. Round $r=R+1$ in a memory experiment represents the final data qubit measurement, which provides the stabilizer information. Note that in a stability experiment, we do not have stabilizer information from the final measurement of the data qubits. For memory experiments, $p_{\rm aux}$ is used to train the NN to only learn from the data gathered in the QEC rounds, excluding the final data qubit measurements. For more information on the architecture for the memory experiment, we refer to Ref.~\cite{Varbanov25}. }
    \label{fig:nn_architecture}
\end{figure}

\begin{table}[t]
\caption{Neural network (NN) sizes used in this work for each of the blocks shown in Fig.~\ref{fig:nn_architecture} and total number of free weights. $n_i$ refers to the number of layers in block $i \in \{\mathrm{LSTM}, \mathrm{Eval}\}$ and $d_i$ the dimension of these layers. The number of free weights depends on the given set of inputs (Table~\ref{tab:nn_inputs}), but the changes are $< 1$k. }
\label{tab:nn_sizes}
\begin{ruledtabular}
\begin{tabular}{lccccc}
Experiment & $n_{\mathrm{LSTM}}$ & $d_{\mathrm{LSTM}}$ & $n_{\mathrm{Eval}}$ & $d_{\mathrm{Eval}}$ & \# weights \\ \colrule
Repetition-5     & 2                   & 24                  & 2                   & 24                  & $\sim$ 9k  \\
Stability-7  & 4                   & 32                  & 2                   & 32                  & $\sim$ 31k \\ 
\end{tabular}
\end{ruledtabular}
\end{table}

\subsubsection{Memory experiment} \label{sec:mem_exp_data}

We use the same NN architecture for the memory experiment as Ref.~\cite{Varbanov25} [Fig.~\ref{fig:nn_architecture}(a)], which is based on the Long Short-Term Memory (LSTM) layers to process an arbitrary number of QEC rounds. This architecture has two outputs: $p_{\rm out}$ and $p_{\rm aux}$, with $p_{\rm aux}$ used to avoid the NN focusing on just the final data qubit information. 

The NN sizes for the Repetition-5 experiments are summarized in Table~\ref{tab:nn_sizes}. Regarding the training hyperparameters, we use a learning rate of $10^{-3}$ and a batch size of 64. The training is stopped early if the validation loss has not decreased over the last 100 epochs or if 500 epochs are reached. These parameter values allow for stable trainings, reaching similar accuracy with different random seeds. 

A Repetition-5 circuit is specified by the initial data qubit bitstring $\Vec{q_d} \in \mathbb{Z}_2^{\otimes 3}$, the total number of QEC rounds $R$, the (random) logical flips at the beginning of each QEC round $\Vec{f} \in \mathbb{Z}_2^{\otimes R}$, the execution ($u=1$) or not ($u=0$) of the LRUs, and the effective leakage rate $\widetilde{L_1} \in \mathbb{R}_+$. Unlike previous repetition-code memory experiments in the literature~\cite{som_Google25, som_Acharya23, som_Ali24}, our experiments include random logical flips to avoid cheating in the decoding task~ (Section~\ref{sec:cheating}). We define a single realization of a circuit as a shot. The training dataset has 20 shots for each circuit specified by $(\Vec{q_d}, R, \Vec{f}, u, \widetilde{L_1}) \in \mathbb{Z}_2^{\otimes 3} \times \mathcal{R}_{\mathrm{train}}^{\mathrm{mem}} \times \mathcal{F}_{\mathrm{train}} \times \mathbb{Z}_2 \times \Tilde{\mathcal{L}}_{\mathrm{mem}}$ with $R \in \mathcal{R}_{\mathrm{train}}^{\mathrm{mem}} = \{1,  2,  3,  4,  5,  6,  7,  8, 11, 14, 17, 20, 23, 26, 29, 32, 35, 38, 41\}$, experimental values $\widetilde{L_1}\in \Tilde{\mathcal{L}}_{\mathrm{mem}} = \{0.98(1)\%$, $1.58(1)\%$, $2.6(1)\%$, $3.63(1)\%$, $4.9(1)\%\}$ and $|\mathcal{F}_{\mathrm{train}}| = 548$. The elements $\Vec{f} \in \mathcal{F}_{\mathrm{train}}$ have been sampled from a uniform distribution fixing $\Vec{f}|_1 = \sum_{i=2}^R \Vec{f}|_i \mod 2$. This choice has been made to enforce $\sum_{i=1}^R \Vec{f}|_i \mod{2}= 0$, so that the ideal logical outcome matches the initial logical state. Note that the first random logical flip can be removed as we prepare all possible bitstrings with the same probability. 
The total number of shots in the training datasets is $\sim 1.7 \cdot 10^7$. The validation dataset contains the same circuits as the training dataset except for $|\mathcal{F}_{\mathrm{val}}| = 44$, leading to a total of $\sim 1.4 \cdot 10^6$ shots. The elements from $\mathcal{F}_{\mathrm{val}}$ have been sampled using the same method as in the training dataset. The testing dataset has 10,730 shots for each circuit $(\Vec{q_d}, R, \Vec{f}, u, \widetilde{L_1}) \in \mathbb{Z}_2^{\otimes 3} \times \mathcal{R}_{\mathrm{test}}^{\mathrm{mem}} \times \{ \Vec{0} \} \times \mathbb{Z}_2 \times \Tilde{\mathcal{L}}_{\mathrm{mem}}$ with $\mathcal{R}_{\mathrm{test}}^{\mathrm{mem}} = \{1,  2,  3,  4,  5,  7,  8, 10, 13, 16, 21, 26, 33, 42, 53, 68, 87, 108\}$, leading to a total of $\sim 1.7 \cdot 10^7$ shots. Note that the testing dataset does not include the random logical flips $\vec{f}$ at the beginning of each QEC round and reaches a higher $R$ compared to the training and validation datasets. The random logical flips are not included because the NNs can only learn to cheat during the training. The testing dataset has experiments with more than 100 rounds to reach $p_L \sim 0.5$ and to show that the NNs are not cheating (Fig.~\ref{fig:nn_cheating}).

\subsubsection{Stability experiment}

We are not aware of previous work decoding stability experiments with NNs. 
Our architecture for the Stability-7 experiments is a slightly modified version of the architecture for memory experiments in Ref.~\cite{Varbanov25} with a single output, $p_{\mathrm{out}}$ [Fig.~\ref{fig:nn_architecture}(b)]. As stability experiments do not have stabilizer information in the final data qubit measurements, we have not included $p_{\rm aux}$ in the NN architecture. We tried different variants of the NN architecture that included encoding layers~\cite{som_Ali24} and/or processed the first round differently than the bulk, but we did not see a performance increase worth the extra training. 

The NN sizes for the Stability-7 experiments are summarized in Table~\ref{tab:nn_sizes}. 
A single LSTM layer does not have long memory or retention~\cite{singh2016multi}, so we use several of them to capture enough time correlations (from measurement errors and leakage events) to achieve good performance in stability experiments. Regarding the training hyperparameters, we use learning rates between $[1 \cdot 10^{-3}, 2\cdot 10^{-3}]$ and a batch size of 128. The training is stopped early if the validation loss has not decreased over the last 100 epochs or if 500 epochs are reached. We had more difficulty achieving stable trainings for the stability experiments compared to memory experiments, which we associate to the larger NN sizes and to the difficulty of achieving very low logical error probabilities at high $R$. 

A Stability-7 circuit is specified by the initial measure qubit bitstring $\Vec{q_a} \in \mathbb{Z}_2^{\otimes 4}$, $R$, the execution of the LRUs at the end of each round, and the effective leakage rate. Unlike previous stability experiments~\cite{som_Caune24}, we initialize the measure qubits in different bitstrings to avoid cheating in the decoding task (Section~\ref{sec:cheating}).
All three datasets (training, validation, and testing) contain the same circuits $(\Vec{q_a}, R, u, \widetilde{L_1}) \in \mathbb{Z}_2^{\otimes 4} \times \mathcal{R}_{\mathrm{train}}^{\mathrm{stab}} \times \mathbb{Z}_2 \times \Tilde{\mathcal{L}}_{\mathrm{stab}}$ with $R \in \mathcal{R}_{\mathrm{train}}^{\mathrm{stab}} = \{r \in \mathbb{N} \; | \; 3 \leq r \leq 40\}$ and experimental values $\widetilde{L_1}\in \Tilde{\mathcal{L}}_{\mathrm{stab}}$ $= \{1.2(1)\%, 2.8(1)\%, 3.7(1)\%,$ $5.3(2)\%,$ $7.6(2)\%\}$. Note that $\mathcal{R}_{\mathrm{train}}^{\mathrm{stab}}$ does not include circuits with one QEC round because there is no logical protection at that point since the input state of the measure qubits $\Vec{q}_a$ is random. For circuits with two QEC rounds there is only error detection with respect to the logical observable, see its definition in Section~\ref{sec:stab-log}, so we exclude these circuits as well. The number of shots per circuit are $\sim 4,100$, $\sim 330$, and 625 for the training, validation, and testing datasets, respectively. This leads to a total number of shots of $\sim 1.3\cdot 10^7$ (training), $\sim 1.0 \cdot 10^6$ (validation), and $\sim 1.9 \cdot 10^6$ (testing).

\subsubsection{Input and training duration} \label{sec:nn_inputs}

The inputs to the NN depend on whether (1) they have access to 3RO or 2RO, 
and whether (2) they are used to decode a Repetition-5 or Stability-7 experiment. The inputs for each case have been summarized in Table~\ref{tab:nn_inputs} and are together, for all rounds $r$, fed to the NN. We have used a similar set of inputs as Refs.~\cite{som_Ali24} and \cite{Bausch24}. Note that the measure qubits are not reset after their measurement in the experiments, thus making the detector definitions non-standard. Using the nomenclature of Table~\ref{tab:nn_inputs}, the detectors for the memory experiment are given by $d_{r=1,a} := m_{r=1, a}$, $d_{r=2,a} := m_{r=2, a}$, $d_{2<r \leq R,a} := m_{r, a} \oplus m_{r-2, a}$, and $d_{r=R+1,a} := s_{a} \oplus m_{R, a} \oplus m_{R-1, a}$, with $s_{a}$ the stabilizer value obtained from the final data qubit measurements~\cite{som_Varbanov20}. For the stability experiment, they are given by $d_{r=2,a} := m_{r=2, a}$ and $d_{2<r \leq R,a} := m_{r, a} \oplus m_{r-2, a}$. Note that there are no detectors associated with the first round in stability experiments (Section~\ref{sec:log-obs}).

\begin{table}[t]
\caption{Input to the NN which depends on the type of experiment that is decoded. The labels for the inputs are: $a$ the measure qubit index, $r$ the QEC round, $\{m_{r, a}\}$ the binary measure qubit outcomes, $\{\Tilde{m}_{r, a}\}$ the ternary measure qubit outcomes, $\{d_{r, a}\}$ the detector information (in the absence of measure qubit resets), and the leakage flags of the measure qubits~\cite{som_Ali24}, i.e. $\{l_{r, a} := 1 \;\mathrm{if}\;\Tilde{m}_{r,a}=2 \;\mathrm{otherwise}\; 0\}$. 
The asterisk (*) indicates that the leakage flags for the data which is used to calculate $z_{\rm raw}$ (Section~\ref{sec:stab-log}), are set to 0 to avoid the possibility of cheating by the NN as it might gain state-dependent information. Note also that no leakage information is provided in the memory experiments for the data qubit measurements at the end, for essentially the same reason~\cite{Varbanov25, Bausch24}.}
\label{tab:nn_inputs}
\begin{ruledtabular}
\begin{tabular}{lcc} 
           & Memory experiment                                      & Stability experiment               \\ \hline
three-level readout (3RO) & $\{\Tilde{m}_{r, a}\}$, $\{d_{r, a}\}$, $\{l_{r, a}\}$ & $\{d_{r, a}\}$, $\{l_{r, a}\}$(*) \\
two-level readout (2RO) & $\{m_{r, a}\}$, $\{d_{r, a}\}$                         & $\{d_{r, a}\}$                   \\ 
\end{tabular}
\end{ruledtabular}
\end{table}

Finally, the training of the NNs for the Repetition-5 experiments took $\sim$ 6~hours and for the Stability-7 experiments $\sim$ 12~hours. The difference is mainly due to the bigger size of the NNs used for the stability experiments (Table~\ref{tab:nn_sizes}).

\subsection{Logical observables and decoding success} \label{sec:log-obs}

The output of the NN used to benchmark its performance is the probability estimate that a logical observable flip has occurred during the given shot of the circuit, labelled $p_{\mathrm{out}} \in [0, 1]$ in Fig.~\ref{fig:nn_architecture}. Note that we also use ensembling of the NN output for enhanced performance (Section~\ref{sec:ens}). The logical observables are defined below for the memory and stability experiments. To determine whether the NN decodes a shot correctly or not, its output $p_{\mathrm{out}}$ is thresholded at $p_{\mathrm{out}}=1/2$, giving the binary value $z_{\mathrm{flip}} \in \mathbb{Z}_2$. This output is used as follows to assess the performance of the NNs.

\subsubsection{Memory experiment}
The logical observable in the memory experiments is simply the logical measurement $z_L$ of the logical qubit after the QEC rounds.
Based on the final data qubit measurements, we compute the uncorrected or raw logical observable outcome, $z_{\mathrm{raw}} \in \mathbb{Z}_2$ as the parity of all the outcomes. Note that this choice is not unique and may affect the performance of the NN as the NN may choose to ignore errors which do not affect $z_{\rm raw}$.
The ideal logical observable outcome, corresponding to the logical input state  $z_{\mathrm{ideal}} \in \mathbb{Z}_2$ is known but not provided to the NN: the NN has no knowledge of this input state due to the initial data qubit randomization $\Vec{q}_d$. The shot is correctly decoded if $z_{\mathrm{flip}} \oplus z_{\mathrm{raw}} = z_{\mathrm{ideal}}$, and incorrectly otherwise. The logical error probability $p_L$ is the probability that the NN incorrectly decodes a shot.

\subsubsection{Stability experiment}
\label{sec:stab-log}
The logical observable in a stability experiment is the product of linearly-dependent parity checks of the code, whose product is (in principle) always $+1$ in the absence of errors~\cite{som_Gidney22}. Due to the different choices for the initial measure qubit states $\Vec{q_a}$, this ideal value ($z_{\rm ideal}\in \mathbb{Z}_2$) is randomized and not known to the NN. We take the raw uncorrected value of the logical observable in a stability experiment, $z_{\rm raw}$, as the product of all measure qubit outcomes at the last odd round. As in a memory experiment, the choice for $z_{\rm raw}$ is not unique. However, one needs to be careful with its definition in the absence of measure resets to ensure that the logical observable information is not present in the detector information. For example, the logical observable cannot be defined as the product of all measure qubit outcomes in the second round, because it will correspond to $\prod_a d_{r=2,a}$ (Section~\ref{sec:nn_inputs}). In fact, the logical observable cannot be defined as the product of all measure qubit outcomes in an even round. Like in the memory experiment, we declare decoding failure when $z_{\rm flip} \oplus z_{\mathrm{raw}} \neq z_{\mathrm{ideal}}$. 

We have observed better logical performance with our specific choice of $z_{\rm raw}$. This can be viewed as a consequence of an imbalanced dataset~\cite{Krawczyk16} when using other definitions for $z_{\rm raw}$, which is a known issue for classification tasks. The true label that the NN needs to predict, $z_{\mathrm{raw}} \oplus z_{\mathrm{ideal}}$, is imbalanced towards 0 for the other definitions, meaning that it is very common to find $z_{\mathrm{raw}} \oplus z_{\mathrm{ideal}} = 0$ in the training dataset. 
For example, if the observable is defined as the product of all measure qubit outcomes from the first round, then the fraction of shots with $z_{\mathrm{raw}} \oplus z_{\mathrm{ideal}} = 0$ is 83\%. This imbalance is caused by the low probability of the observable being flipped in the first rounds, because it is unlikely that, e.g., a measurement error occurred. 
Therefore, in such case, the NN basically has an easier task and thus it does not learn very well.
We emphasize that the logical performance never depends on the raw observable definition once the model has been trained, i.e., once the weights of the NN are fixed. Of course this also holds for memory experiments or lattice surgery experiments. The reason is that all raw logical observables are equivalent up to XORs with detectors which can be viewed as (space-time) stabilizers. For example, for the memory experiment on the bit-flip repetition code, one can get $z_{\rm raw}$ from any of the single-qubit (data qubit) measurements, as the logical $z_L$ is any single-qubit $z$.

\subsection{Ensembling}
\label{sec:ens}
Ensembling is a machine learning technique where one trains several NNs, which form an ensemble, and averages their outputs, $\{p_{\mathrm{out}, i}\}$, to obtain a more accurate prediction, $\Tilde{p}_{\mathrm{out}}$. 
Each of our ensembles consists of $n=5$ NNs that are trained with the same datasets and hyperparameters. 
We use the geometric mean to average their outputs, 
$\Tilde{p}_{\mathrm{out}} = \sqrt[n]{\prod_{i=1}^n p_{\mathrm{out}, i}}$, but we obtain similar performance when using the standard average, $\Tilde{p}_{\mathrm{out}} = \sum_{i=1}^n p_{\mathrm{out}, i} / n$. The averaged output $\Tilde{p}_{\mathrm{out}} \in [0, 1]$ corresponds to an improved probability estimate of a flip in the logical observable and is processed as described in Section~\ref{sec:NN-para}. 

\subsection{Increasing leakage in the training dataset} \label{sec:increased_leakage_in_training}

\begin{figure*}[t]
    \centering
    \includegraphics[width=\textwidth]{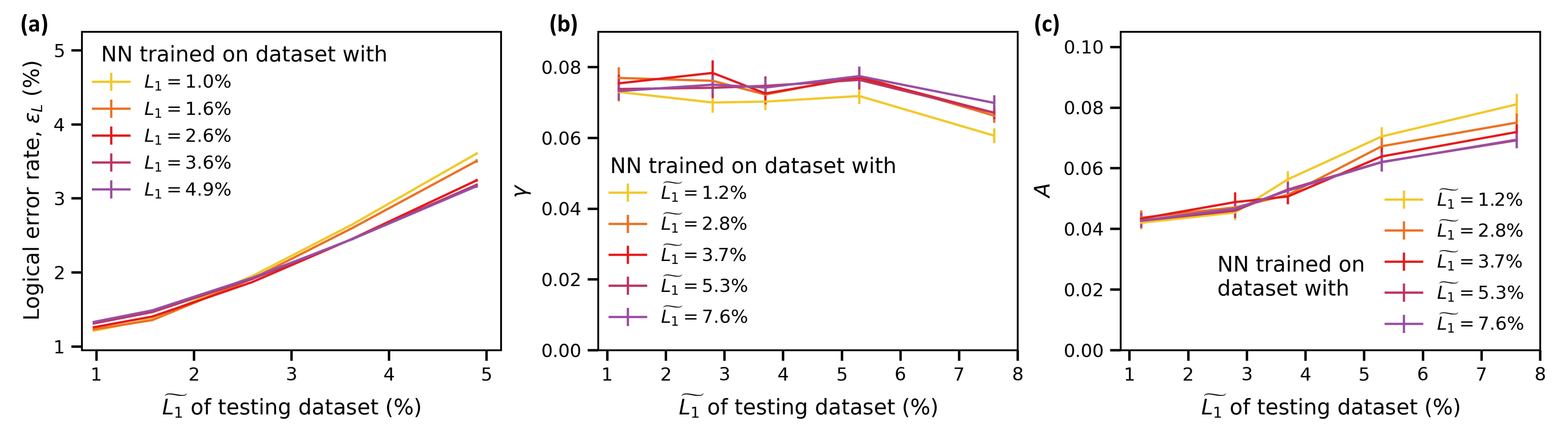}
    \caption{Neural network (NN) ensemble decoder performance for (a) the Repetition-5 experiment and (b, c) the Stability-7 experiment with LRUs and 3RO, when evaluated using testing datasets with different effective leakage rates, $\Tilde{L}^{\mathrm{test}}_1$. Even though the NNs were trained for different leakage rates, they all have similar performance for a given $\Tilde{L}^{\mathrm{test}}_1$. For memory experiments, the (overall) best-performing NN for each $\Tilde{L}^{\mathrm{test}}_1$ corresponds to the one trained with such $\Tilde{L}^{\mathrm{test}}_1$, as expected. For stability experiments, identifying the best-performing NNs is more complicated because there are two metrics (the parameters $\gamma$ and $A$) and they both have larger error bars. }
    \label{fig:increased_leakage_in_training}
\end{figure*}

Training the NNs with datasets that have larger error rates than the testing dataset has been shown to improve their logical accuracy~\cite{Peters25, Varbanov25, Chamberland18, Liu19}. We have tried this strategy in both the memory and stability experiments by decoding each test dataset with the NNs trained for other leakage rates, i.e., $\Tilde{L}^{\mathrm{train}}_1 \neq \Tilde{L}^{\mathrm{test}}_1$. 
We do not observe any clear improvement in the logical performance (Fig.~\ref{fig:increased_leakage_in_training}). Nevertheless, the results show that NNs achieve similar performance when they are trained with a dataset that has a different leakage rate than the actual experiment. Such resilience to noise fluctuations can be useful to avoid retraining the NNs every time there is a change in the noise of the quantum device. In particular, for the Repetition-5 experiment, at low leakage rates, the NNs trained with 3RO have very similar performance among them. At higher leakage rates, the NNs trained with 2RO have more similar performance between them, as compared to the NNs trained with 3RO. 
Regarding the Stability-7 experiment, the NNs trained with the baseline noise are less accurate at higher leakage rates than the rest, which have similar performance. This trend is observed for all combinations of 2RO or 3RO with or without LRU. 

\section{Cheating modes of the NN}
\label{sec:cheating}
When training a NN to learn how to process detector data to interpret logical observable measurements, care has to be taken to train it in a fashion showing performance that is extendable. That is, one demonstrates NN performance which can be expected to generalize if the same NNs were to be used on (1) an unknown input logical state, (2) for a genuinely quantum code such as the surface code (instead of a repetition code), and (3) for other noise models, for example, state-independent depolarizing noise. The issue is that NNs may more easily decode simpler experiments for a purely classical code in which state-dependent information becomes available to the NN through detector information. 
An example is running a memory experiment for a repetition code without applying echo $X$ gates between each QEC round: due to state-dependent $T_1$ noise, most qubits will decay to $\ket{0}$ after some time, leading to low defect rates and easier decoding. However, for the surface code, $T_1$ noise would not lower the bar for decoding, hence the use of deterministic toggling between logical states at each QEC round in the repetition code to make results extendable. 

As described in Section~\ref{sec:log-obs}, the NNs are trained to predict wether the logical observable has been flipped or not ($z_{\rm flip}$), instead of predicting the corrected logical observable ($z_{\rm flip} \oplus z_{\rm raw}$). Not showing any logical state information to the NN, i.e. not providing it $z_{\rm raw}$, causes its performance to be more extendable to unknown input logical states. This choice is also important for making the results from repetition codes extendible to quantum codes. 
In particular, it makes it more difficult for the NN to use state-dependent information to infer the correct label instead of solving the decoding problem, which is considered as cheating and would not be possible in, for example, the surface code. One also randomizes the initial data qubit bitstring $\Vec{q}_d$ and the initial logical state in memory experiments (thus randomizing $z_{\rm ideal}$) to reduce the possibility of cheating. Similarly, one should randomize $z_{\rm ideal}$ in a stability experiment by randomly picking $\vec{q_a}$. However, in the memory experiment on the repetition code, other cheating modes which use state-dependent information are possible, which we discuss now. 

\begin{figure}[t]
    \centering
    \includegraphics[width=\columnwidth]{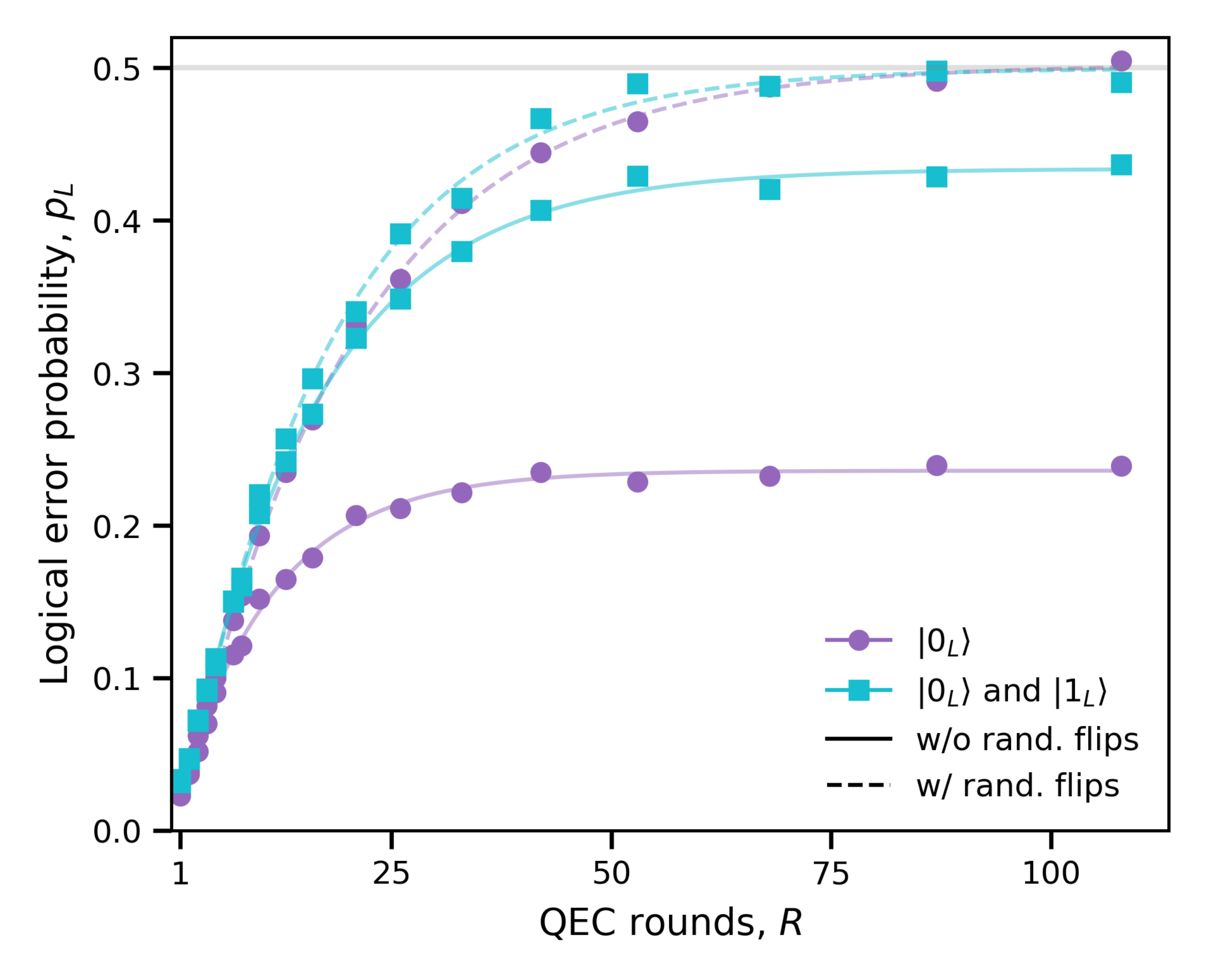}
    \caption{Logical error probability ($p_L$) as a function of the number of rounds ($R$) for the Repetition-5 experiment with and without the random logical flips. The data for ``$\ket{0}_L$'' corresponds to the bistrings 000, 011, 101, and 110; and the data for ``$\ket{0}_L$ and $\ket{1}_L$'' corresponds to the bitstrings 010 and 101. The decoder is a single NN, without ensembling. The fitting formula is $p_L(R) = p_L^{\infty} - \frac{1}{2}(1 - 2\varepsilon_L)^{R+R_0}$, with $p_L^{\infty}$ the plateau of $p_L$ at $R \rightarrow \infty$.}
    \label{fig:nn_cheating}
\end{figure}

For the first cheating observation, we only offer data for $\ket{0}_L$ in all its representative bitstrings to the NN in a memory experiment. For the second cheating observation, we offer data for both $\ket{0}_L$ and $\ket{1}_L$, but with only one representative bitstring for each logical state. The logical performance for the two observations is presented in Fig.~\ref{fig:nn_cheating}, showing that the $p_L$ does not decay to 1/2 as one would expect, but rather to 0.24(1) and 0.43(1), respectively. This indicates that the NN can still use state-dependent information to cheat. We have found that the measure qubit leakage data contains state-dependent information that can reveal the data qubit states, and thus the logical state. 
The main measure qubit leakage source in Fig.~\ref{fig:nn_cheating} is the $\ket{1}_{\rm data}\ket{1}_{\rm meas} \leftrightarrow \ket{0}_{\rm data}\ket{2}_{\rm meas}$ interaction during the CZ gates, thus the leakage events ($\ket{2}_{\rm meas}$) can be associated with a data qubit state changing from $\ket{1}$ to $\ket{0}$ through the CZ interaction. 
From this information, it is possible to know all data qubit states at the beginning and end of the memory experiment and infer both $z_{\rm ideal}$ and $z_{\rm raw}$. Therefore, the NN can obtain the correct $z_{\rm flip}$ by $z_{\rm ideal} \oplus z_{\rm raw}$ without decoding the detector data, thus obtaining a constant logical performance that does not depend on $R$ (Fig.~\ref{fig:nn_cheating}). 

To address the cheating problem in repetition-code memory experiments, we include random logical flips between each QEC round that are not known to the NN, hence the vector $\Vec{f}$ in Section~\ref{sec:mem_exp_data}. These random flips do not change the detector values but scramble the logical information at each round, which makes the NN unable to learn the logical state (Fig.~\ref{fig:nn_cheating}). Performing these random flips is more stringent than randomly preparing different bitstrings and logical states, because these latter ones only scramble logical information per circuit.

\section{Post-selection analysis}
A common strategy to mitigate the impact of leakage is to apply post-selection, which we analyze here for comparison with our LRU protocol. The most straightforward post-selection on leakage involves discarding any experimental shot where a 2 (leakage) outcome is measured on any qubit in any round. Furthermore, since the NN decoder is used, a more sophisticated post-selection method leverages the NN decoder, discarding shots where the prediction confidence is low, as this indicates complex error patterns potentially involving leakage. For a fair comparison, the confidence threshold is set to discard the same number of shots as the post-selection method based on leakage.

\begin{figure}[t]
\includegraphics[width=\columnwidth]{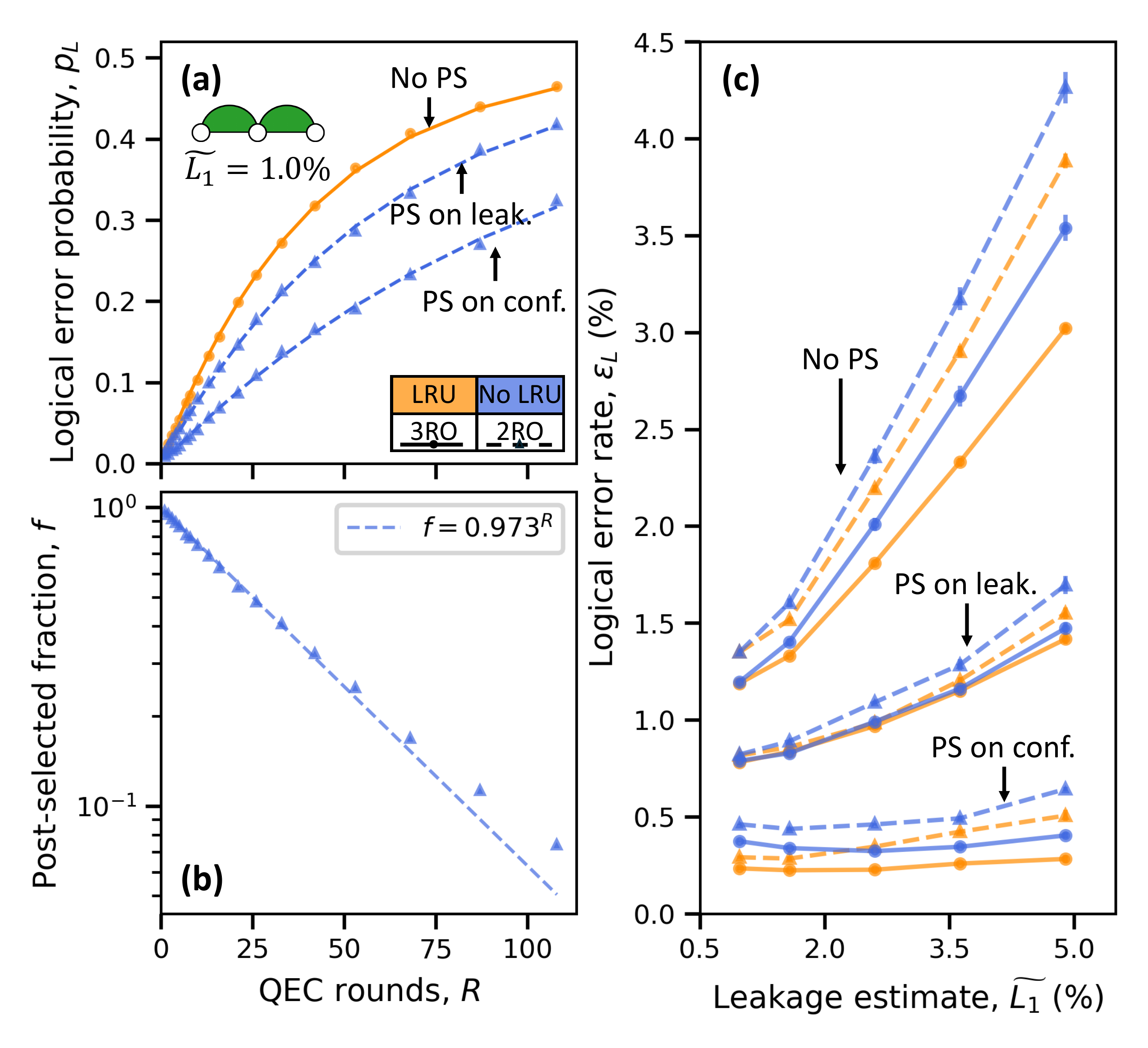}
\caption{Post-selection analysis of memory experiments.
(a) Logical error probability ($p_L$) versus QEC rounds ($R$) at baseline leakage ($\widetilde{L_1} = 1.0\%$). Performance is shown for the 3RO with protocol without post-selection (No PS), 2RO without LRU with post-selection on leakage (PS on leak.), and 2RO without LRU with post-selection on NN decoder confidence (PS on conf.). 
(b) Fraction of experimental shots kept ($f$) as a function of $R$ for post-selection. Data is fitted to an exponential decay (dashed line).
(c) Extracted logical error rate ($\epsilon_L$) versus leakage estimate ($\widetilde{L_1}$) for all protocols. The hierarchy of No PS > PS on leak. > PS on conf. holds across all leakage rates. Fits are performed only on rounds with more than 250 shots.
}
\label{ps_memory}
\end{figure}

Figure~\ref{ps_memory}(a) compares the two post-selection strategies to the best-performing protocol (LRU with 3RO) at the baseline leakage rate for the memory experiment. While all methods improve upon the 2RO without LRU without post-selection, both post-selection strategies yield a lower $p_L$ than the 3RO with LRU without post-selection. This is because post-selection purifies the dataset by removing hard-to-correct errors entirely, whereas the 3RO with LRU converts leakage errors into Pauli errors with flags, which are still subject to decoding failures. Post-selection on NN confidence, which uses more information, performs better than simple leakage post-selection. Despite the improvement on logical performance, the high cost of post-selection is evident. The fraction of shots kept, $f$, decays exponentially with $R$ [Fig.~\ref{ps_memory}(b)]. This is why post-selection is not extendable to large-scale experiments. 

The full comparison of logical error rates in Fig.~\ref{ps_memory}(c) confirms that post-selection on NN confidence consistently outperforms simple post-selection on leakage. (Due to the exponential loss of data, fits are performed only on rounds with more than 250 shots remaining.) Notably, the logical error rate for post-selection on leakage still increases with the physical leakage rate. This is likely due to imperfect leakage detection (e.g., readout errors mapping 2 to 1, or leakage to higher states that are not classified) allowing some leakage errors to survive the filter. In contrast, post-selection on NN confidence maintains a nearly flat error rate. Note that post-selection on confidence is distinct from trivially post-selecting on the final logical error rate (i.e., discarding shots because they failed). Our method filters shots based on the ambiguity of the syndrome data as assessed by the decoder, not on the final correctness of the outcome. An ideal, perfectly unbiased decoder would only be unconfident on syndromes that truly lead to a logical error. However, a real-world decoder is imperfect: it can be over-confident in a wrong prediction (due to decoder bias) or under-confident in a correct one. Thus, post-selecting on confidence is a filter on decoder ambiguity, which is a correlated but distinct metric from the true logical error rate.

\begin{figure}[t]
\includegraphics[width=\columnwidth]{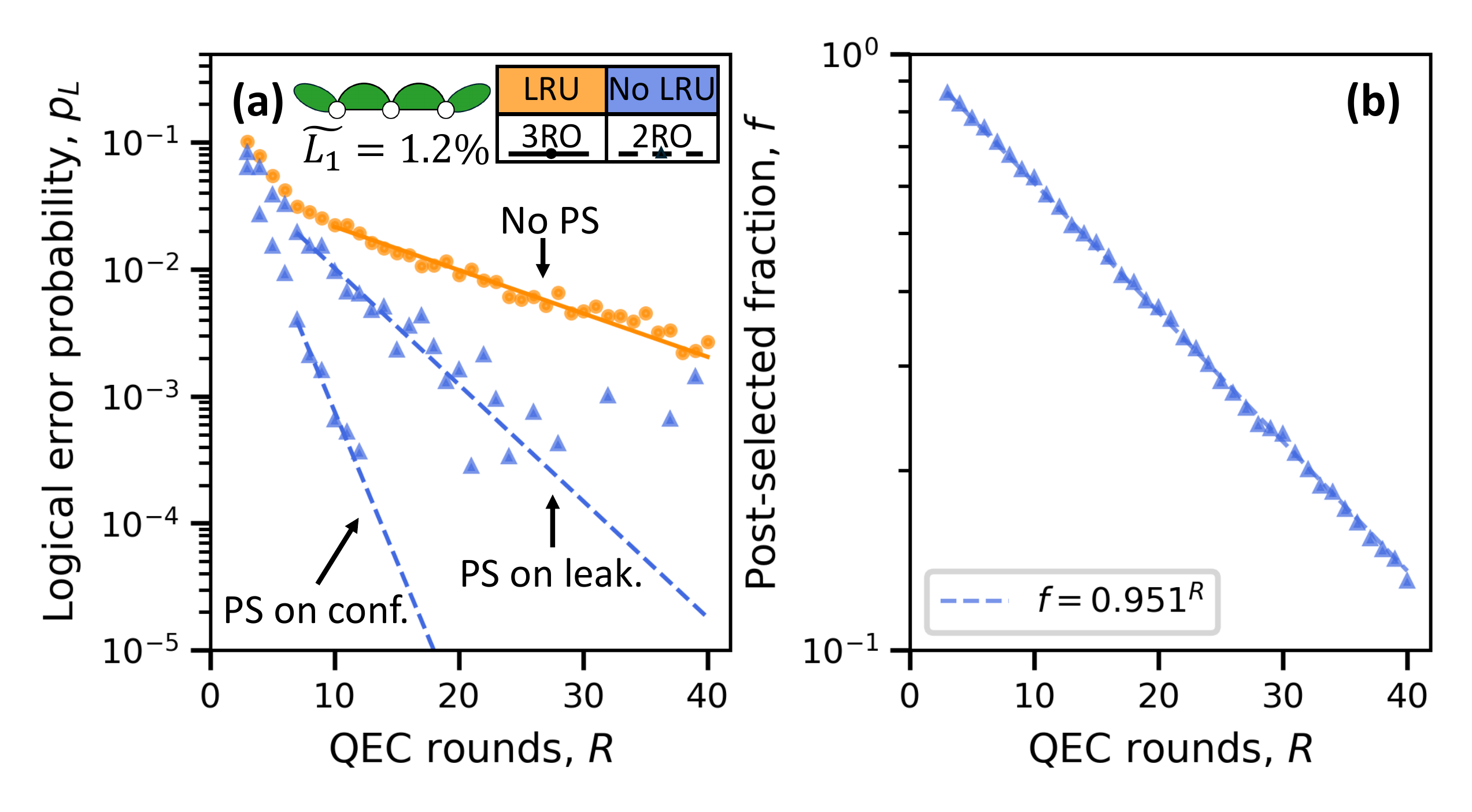}
\caption{Post-selection analysis of stability experiments.
(a) Logical error probability ($p_L$) versus QEC rounds ($R$) at baseline leakage ($\widetilde{L_1} = 1.2\%$). Performance is shown for the 3RO with LRU protocol without post-selection (No PS), 2RO without LRU with post-selection on leakage (PS on leak.), and 2RO without LRU with post-selection on NN decoder confidence (PS on conf.). 
(b) Fraction of experimental shots kept ($f$) as a function of $R$ for post-selection. Data is fitted to an exponential decay (dashed line).
}
\label{ps_stability}
\end{figure}

The same post-selection analysis is applied to the stability experiment. The comparison [Fig.~\ref{ps_stability}(a)] shows the same hierarchy of performance as the memory experiment: post-selection on confidence provides the lowest error rate, followed by post-selection on leakage. The non-exponential curvature in the first few rounds (see main text) is present both with post-selection and without post-selection on leakage. While the effect is less pronounced after post-selecting on leakage, its persistence supports the conclusion from Fig.~\ref{ps_memory}(c) that the post-selection on leakage is an imperfect filter. Due to the much lower logical error rates in the stability experiment compared to the memory experiment, there are insufficient statistics to compare all combination of protocols over leakage rates. 

\section{Simulation of QEC experiments}
To better understand the experimental results and the impact of various parameters, we perform numerical simulations of both the memory and stability experiments.

\subsection{Time evolution and error model}
The time evolution of the system is given by solving the master equation using the QuTiP Monte Carlo solver~\cite{Johansson12}, with collapse operators as defined in Ref.~\cite{som_Varbanov20}. Single-qubit gates are treated as ideal operations, subject only to decoherence channels. The CZ gate is modeled as~\cite{som_Varbanov20}:
\begin{equation}
\begin{gathered}
|11\rangle \rightarrow \sqrt{1-4L_1}|11\rangle + e^{\mathrm{i}\phi}\sqrt{4L_1}|02\rangle, \\
|02\rangle \rightarrow \sqrt{1-4L_1}|02\rangle - e^{-\mathrm{i}\phi}\sqrt{4L_1}|11\rangle.
\end{gathered}
\label{cz model}
\end{equation}
As discussed in Ref.~\cite{som_Varbanov20}, the phase $\phi$ does not have a discernible effect on the logical error rate. Our simulation also confirms this observation, and $\phi$ is thus fixed at $\pi/2$. Besides, the system is subject to decoherence during the CZ gates, but any increased dephasing is not considered in this model for simplicity. The measurement process is modeled as a simplified three-step sequence. First, an ideal state projection is applied. This is followed by a classical assignment, which is quantified by the assignment matrix [see Fig.~2(c--d)]. Finally, a state transition is applied, quantified by the population transfer matrix [see Fig.~2(e--f)]. During the readout of measure qubits, data qubits are modeled as idling and are subject to decoherence, with $X$ gates applied in the middle. While a more complex, non-sequential model using a readout butterfly~\cite{som_Marques23} could more accurately capture the full measurement dynamics, we use this sequential model for simplicity.

\subsection{Results}
\begin{figure*}[t]
\includegraphics[width=\textwidth]{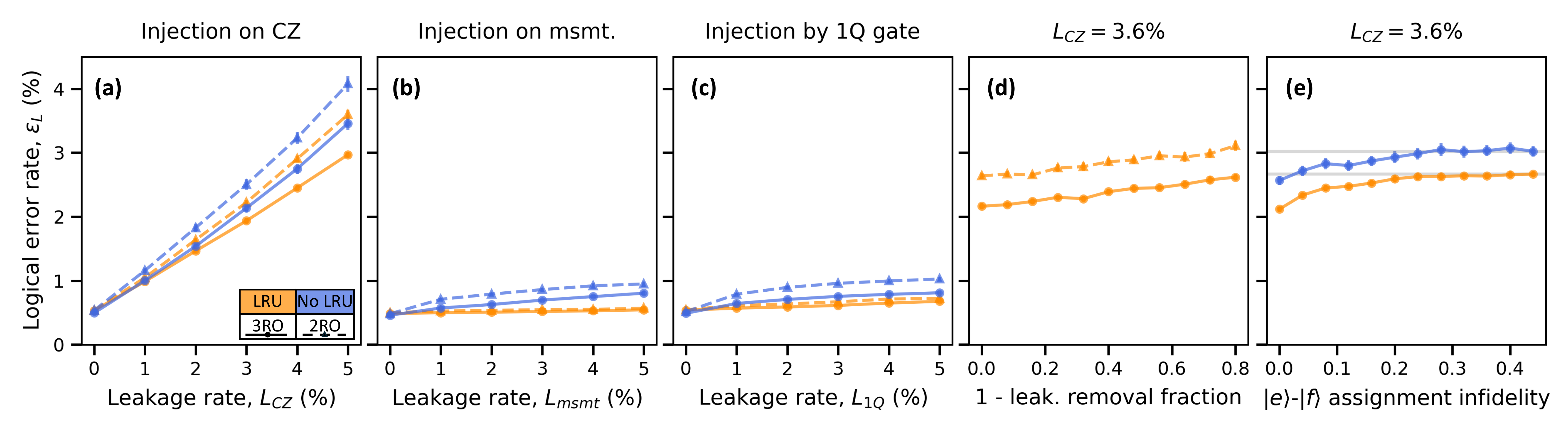}
\caption{Simulation of the memory experiment.
(a) Logical error rate ($\epsilon_L$) versus leakage $L_{CZ}$ injected on CZ gates. The result shows good agreement with experimental data in Fig. 4(e) in the main text.
(b) $\epsilon_L$ versus $L_{msmt}$ injected during readout of measure qubits. Here $L_{msmt}$ is defined as $P(f|e)/4$.
(c) $\epsilon_L$ versus $L_{1q}$ injected via single-qubit $R_X^{12}(\theta)$ gates. Here $L_{1Q}$ is defined as $(\mathrm{sin}^2(\theta/2))/4$.
(d) $\epsilon_L$ versus residual $|f\rangle$ population, at a fixed intermediate CZ leakage rate of $L_1=3.6\%$ with LRU.
(e) $\epsilon_L$ versus $\kete$-$\ketf$ assignment infidelity, at a fixed intermediate CZ leakage rate of $L_1=3.6\%$ with 3RO.
The gray lines in (e) are guides for the eye indicating the endpoint of each curve. The decoder was trained on 96,000 shots and evaluated on 12,000 shots for each point. Each shot in the training dataset has an individual random logical bit flip configuration.
}
\label{memory_sweep_sim}
\end{figure*}

\begin{figure*}[t]
\includegraphics[width=\textwidth]{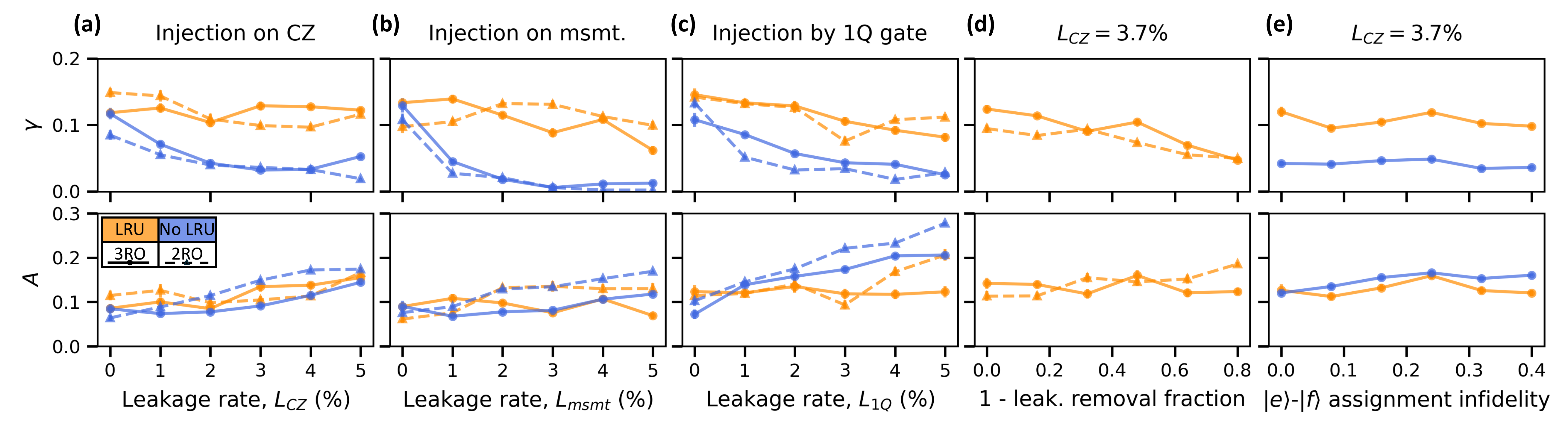}
\caption{Simulation of the stability experiment. 
(a) $\gamma$ and $A$ versus leakage $L_{CZ}$ injected on CZ gates.
(b) $\gamma$ and $A$ versus leakage $L_{msmt}$ injected during readout of measure qubits. 
(c) $\gamma$ and $A$ versus leakage $L_{1Q}$ injected via single-qubit $R_X^{12}(\theta)$ gates.
(d) $\gamma$ and $A$ versus (1 - leakage removal fraction), at a fixed intermediate CZ leakage rate of $L_{CZ}=3.7\%$ with LRU.
(e) $\epsilon_L$ versus $\kete$-$\ketf$ assignment infidelity, at a fixed intermediate CZ leakage rate of $L_1=3.7\%$ with 3RO.
Due to the fact that this simulation is time-consuming, the training and evaluation datasets are structured differently from the experimental ones. The decoder uses 16,000 individual shots for training, with rounds $R$ from 3 to 28 in increments of 3, and 16,000 individual shots at $R=28$ for evaluation. For both datasets, the data at a given  $R$ includes the results extracted from all individual shots that run for number of rounds greater than or equal to $R$. This partially compensates underfitting issues in training due to the lower shot count.
}
\label{stability_sweep_sim}
\end{figure*}

We conduct five distinct simulation sweeps for both the memory (Fig.~\ref{memory_sweep_sim}) and stability (Fig.~\ref{stability_sweep_sim}) experiments to isolate the effects of different parameters. Results are decoded by the same NN decoders as in experiment. First, to investigate the impact of various leakage sources [Fig.~\ref{memory_sweep_sim}(a--c) and ~\ref{stability_sweep_sim}(a--c)], one of them is swept at a time, keeping the other leakage sources at zero. Other noise sources use parameters extracted from experiment. The CZ gate leakage is swept by varying $L_1$ in the gate model (Eq.~\ref{cz model}). The measurement-induced leakage is swept by varying the probability $P(f|e)$ in the population transfer matrix, which represents the excitation from $\kete$ to $\ketf$. Besides, it is common to insert an $R_X^{12}(\theta)$ gate on each measure qubit after it is prepared to the superposition states at each round~\cite{som_Mcewen21, som_Yang24}. This injection is also compared and swept by varying the rotation angle $\theta$. Second, we investigate the trade-off between the leakage removal fraction and $\ketf$ assignment infidelity, as discussed in the main text (Fig. 2). For these sweeps, these two parameters are varied independently to observe their effect on logical performance [Fig.~\ref{memory_sweep_sim}(d--e) and Fig.~\ref{stability_sweep_sim}(d--e)], while other error sources use parameters extracted from experiment with an intermediate CZ leakage level. The leakage removal fraction is swept by varying $P(f|f)$ in the population transfer matrix (which consequently adjusts $P(e|f)$). The $\ketf$ state assignment infidelity is swept by varying $P(1|f)$ and $P(2|e)$ in the assignment matrix (which determines $P(2|f)$ and $P(1|e)$ accordingly).

For the memory experiment, the simulations reveal that not all leakage sources are equally detrimental; leakage from CZ gates is a far more significant error source than single-qubit leakage. We first validate our simulation model against the CZ gate leakage injection [Fig.~\ref{memory_sweep_sim}(a)], which shows a good match with experiment [Fig. 4(e)]. Next, the simulation shows that $\epsilon_L$ increases much more slowly with measurement-induced leakage compared to CZ gate leakage. This is because this leakage source only affects the measure qubits and does not introduce errors on the data qubits (leakage mobility~\cite{som_Varbanov20, som_Camps24} not included in the simulation). The logical error rate is then upper-bounded by a simple majority vote on the final data qubit measurements, which ignores all measure qubit information. However, a leakage on a CZ gate is associated with a bit flip on data qubits, as the transition is between $|11\rangle$ and $|02\rangle$. This spatially correlated error is more difficult to correct for the decoder. Notably, with the LRU enabled, $\epsilon_L$ remains almost constant over measurement-induced leakage rates, demonstrating that the LRU effectively mitigates the impact of this leakage source even without resetting the measure qubits each round. Similarly, sweeping injected leakage via $R_X^{12}(\theta)$ gates shows similar behavior [Fig.~\ref{memory_sweep_sim}(c)], as it also represents a single-qubit leakage on measure qubits. In all three sweeps, the combination of LRU and 3RO consistently provides the best performance.

The simulations also reveal that the logical performance is more sensitive to the $\kete$-$\ketf$ assignment fidelity than to the leakage removal fraction. At an intermediate CZ gate leakage level $L_1$=3.6\%, sweeping the residual $\ketf$ population shows that the logical error rate increases only approximately linearly [Fig.~\ref{memory_sweep_sim}(d)]. We expect the first few points of the curve to be even flatter at lower leakage level. In contrast, sweeping the $\kete$-$\ketf$ assignment infidelity causes the logical error rate to increase much more rapidly before saturating [Fig.~\ref{memory_sweep_sim}(e)], indicating that assignment fidelity is the more critical parameter at low error rates. It also shows that the 3RO provides a benefit only if the $\kete$-$\ketf$ assignment fidelity is $\gtrsim$ 80\%. This finding guides our experimental choice in Fig. 2.

For the stability experiment, a general observation across all simulations [Fig.~\ref{stability_sweep_sim}(a--e)] is that the simulation overestimates the logical performance compared to the experimental results. This discrepancy may be due to error sources that are not captured by the model, such as leakage to higher excited states, performance fluctuations over time, crosstalk, or residual ZZ couplings. Uncertainty is higher than in experiment due to limited number of shots. Still, general trends can be extracted from the simulation results.

In stark contrast to the memory experiment, measurement-induced leakage [Fig.~\ref{stability_sweep_sim}(b)] is now the most damaging leakage source among the three sources [Fig.~\ref{stability_sweep_sim}(a--c)]. The error suppression factor $\gamma$ decays rapidly to almost zero at high measurement leakage rates. This is because, in the stability experiment, the logical observable is defined by the measure qubit readout outcomes themselves. Therefore, errors on the readout of measure qubits, especially correlated leakage errors, are devastating to the logical performance. Errors on data qubits [from CZ gate leakage events, Fig.~\ref{stability_sweep_sim}(a)] has a much smaller impact. The effect of injected $X_{12}$ gates [Fig.~\ref{stability_sweep_sim}(c)] is similar to, but less severe than, direct measurement-induced leakage.

Still contrary to the memory experiment, $\gamma$ is more sensitive to the leakage removal fraction than the $\kete$-$\ketf$ assignment fidelity. Further study of comparing the performance between them in lattice surgery (which combines the memory experiment and the stability experiment) will be valuable.

\FloatBarrier

\end{document}